\documentclass[onecolumn,secnumarabic,nobibnotes,superscriptaddress,aps,prx]{revtex4-2}
\usepackage{color, xcolor, colortbl}
\usepackage{graphicx}
\usepackage{geometry}
\usepackage{amsthm,amsmath,amssymb,amsfonts}
\usepackage{dcolumn}
\usepackage{url}
\usepackage{epstopdf}
\usepackage{enumitem}
\usepackage{algorithm}
\usepackage{algpseudocode}
\usepackage{bm}
\usepackage[caption=false]{subfig}
\usepackage{appendix}
\usepackage{multirow}
\usepackage{booktabs}
\usepackage{braket}
\usepackage[english]{babel}
\usepackage{hyperref}
\usepackage[capitalize]{cleveref}
\usepackage{xpatch}
\usepackage{tikz}
\usepackage{adjustbox}
\usepackage{xspace}
\usetikzlibrary{quantikz}

\newcommand{\Tr}{\operatorname{Tr}}

\newcommand{\mc}[1]{\mathcal{#1}}

\newcommand{\wt}[1]{\widetilde{#1}}

\newcommand{\norm}[1]{\left\lVert#1\right\rVert}

\newcommand{\ud}{\,\mathrm{d}}
\newcommand{\Or}{\mathcal{O}}
\newcommand{\EE}{\mathbb{E}}

\providecommand{\definitionname}{Definition}
\providecommand{\assumptionname}{Assumption}
\providecommand{\corollaryname}{Corollary}
\providecommand{\lemmaname}{Lemma}
\providecommand{\propositionname}{Proposition}
\providecommand{\remarkname}{Remark}
\providecommand{\theoremname}{Theorem}
\usepackage{bbm}

\makeatletter
\newenvironment{breakablealgorithm}
  {
   \begin{center}
     \refstepcounter{algorithm}
     \hrule height.8pt depth0pt \kern2pt
     \renewcommand{\caption}[2][\relax]{
       {\raggedright\textbf{\fname@algorithm~\thealgorithm} ##2\par}%
       \ifx\relax##1\relax 
         \addcontentsline{loa}{algorithm}{\protect\numberline{\thealgorithm}##2}%
       \else 
         \addcontentsline{loa}{algorithm}{\protect\numberline{\thealgorithm}##1}%
       \fi
       \kern2pt\hrule\kern2pt
     }
  }{
     \kern2pt\hrule\relax
   \end{center}
  }
\makeatother

\usetikzlibrary{fit}
\tikzset{%
  highlight/.style={rectangle,rounded corners,fill=blue!15,draw,fill opacity=0.3,thick,inner sep=0pt}
}

\begin{document}

\title{A Theory of Finite-Noise Optima and Generalization in Quantum Machine Learning}
\author{Ziyu Zhang}
\thanks{These authors contributed equally.}
\affiliation{Department of Chemistry, University of Washington, Seattle, Washington 98195, USA}
\author{Zikang Jia}
\thanks{These authors contributed equally.}
\affiliation{Department of Mathematics, University of Michigan, Ann Arbor, Michigan 48109, USA}
\author{Xiaosong Li}
\affiliation{Department of Chemistry, University of Washington, Seattle, Washington 98195, USA}
\author{Yulong Dong}
\email[Corresponding author: ]{dongyl@umich.edu}
\affiliation{Department of Electrical Engineering and Computer Science, University of Michigan, Ann Arbor, Michigan 48109, USA}
\date{\today}

\begin{abstract}
Quantum noise is expected to degrade quantum machine learning by driving circuits away from their noiseless implementations. Yet recent studies show moderate noise can reduce testing error, a behavior unexplained by weak-noise perturbative error accumulation or strong-noise trainability collapse. Here we develop a statistical learning theory connecting microscopic noise processes to macroscopic learning performance. At its heart is a noise-order purity parameter, derived from a surrogate model analysis, that predicts the noise-induced reduction in model complexity and the consequent reduction in the generalization gap. Noise simultaneously increases prediction bias. Their competition explains the intermediate-noise regime left open between these limits. It produces a finite-noise optimum whose location depends on the learning setup and can disappear in the large-sample limit. Numerical experiments validate these predictions. Noise programming can move a model towards this optimum. These results make the non-monotonic effect of noise predictable and provide a route to harness it.
\end{abstract}

\maketitle

\section{Introduction}\label{sec:intro}

Quantum noise is conventionally treated as an error because it drives a physical implementation away from its prescribed computation~\cite{Preskill18_79,Coles21_625,Bharti22_015004}. This view is natural for quantum algorithms whose accuracy relies on a carefully synthesized structure, and it is reinforced in quantum machine learning by two well-studied limits. Near the noiseless limit, analysis and mitigation are organized around deviations from the ideal circuit~\cite{Temme17_180509,Endo21_032001,Cai23_045005,Franca21_1221,Tsubouchi23_210601,Takagi23_210602,Quek24_1648}. At sufficiently large noise and circuit depth, parameter influence is suppressed, training collapses and a deep circuit becomes effectively shallow~\cite{Du21_040337,Coles21_6961,Eisert26_751}. Both limits are destructive, inviting the conclusion that noise must degrade a quantum learning model throughout the range between them. Yet neither limit determines what happens before the onset of the high-noise collapse.

A learned predictor is not required to reproduce a prescribed noiseless computation~\cite{Biamonte17_195,Cerezo22_567}. It adapts jointly to its trainable components, finite data and the physical noise acting on the circuit, and its objective is performance on unseen data rather than circuit fidelity~\cite{Abbas21_403,Caro22_4919,Huang21_2631,Peters23_1210}. Noise can therefore remove useful signal, but it can also suppress excess freedom used to fit the training sample~\cite{Bu22_062431,Bu23_025013,Heyraud22_052421}. The use of stochastic perturbations in training as a source of regularization has been studied in classical machine learning~\cite{Bishop95_108,An96_643,Srivastava14_1929}. Recent quantum studies suggest similar possibilities. Moderate noise can improve variational quantum algorithms and quantum neural networks~\cite{Liu21_023153,Borondo23_8790,Kattemolle23_042426,Eisert25_052441,Sannia24_81,Melnikov25_e00603,Li26_260113275}. Quantum models can also remain robust to noise~\cite{Nguyen20_1,Beer20_808}, while noise redistributes locally important parameter directions~\cite{Lucchi26_045015}. These observations challenge a monotonic picture of noise, but do not yet explain when the improvement occurs, what controls the optimal noise level or why the benefit eventually disappears.

The obstacle is not a lack of bounds, but a mismatch between their mathematical objects and the learning question~\cite{CaroDatta20_14,Bu22_062431,Gyurik23_893,Caro23_3751,GilFuster24_2277}. Norm-based perturbative bounds control worst-case deviation; the triangle inequalities that make them general also discard cancellations and interactions among stochastic components, leaving them valid but uninformative in the intermediate regime~\cite{Temme17_180509,Endo21_032001,Cai23_045005}. Strong-noise analyses instead characterize the loss of representation and trainability~\cite{Franca21_1221,Du21_040337,Coles21_6961,Larocca25_174,Eisert26_751}. Quantum Fisher information describes local state-space geometry and data-dependent learnable directions, but does not directly determine target-dependent testing loss~\cite{Meyer21_539,Abbas21_403,HaugKim24_050603,Lucchi26_045015}. What is missing is a task-aligned analysis that retains the structure of the noisy predictor and follows how its changing statistical complexity affects generalization.

In this work, we fill this gap by developing a statistical learning theory for noisy quantum learning models. Using a noise-order surrogate, we organize the noisy response by the number of noise occurrences and introduce the noise-order purity, which measures how broadly the response is mixed across noise orders. We show that this single quantity controls the effective model complexity and therefore the generalization gap. Noise simultaneously attenuates the learnable signal. The competition between these effects produces a finite-noise optimum whose location is determined by the complete learning setup. Numerical results verify the predicted links from noise-order purity to model complexity and from model complexity to the generalization gap. They further show that the improvement occurs while the model remains trainable, before the strong-noise collapse. This theory complements the established weak- and strong-noise limits by resolving the learning behavior between them. Making this regime predictable also allows noise to be harnessed: controlled perturbations can move a model towards its finite-noise optimum. The same idea may inform how stochasticity is introduced during training and how protection is allocated in partially fault-tolerant architectures.


\section{Results}

\subsection{Finite noise induces non-monotonic generalization}

We first study the finite-noise performance of a noisy QML model on the diabetes regression dataset, which contains baseline clinical measurements from 442 patients and a quantitative measure of disease progression one year later~\cite{efron2004least,sklearn_diabetes}. We use a four-qubit parametrized circuit to predict this outcome from two baseline clinical measurements. We fix the dataset split, circuit architecture, and training protocol throughout the experiment, while varying the physical noise rate $\beta$. For each value of $\beta$, the model is trained and evaluated under the same noisy circuit dynamics. We consider depolarizing, bit-flip, phase-damping, and amplitude-damping noise. The first three channels are unital, whereas amplitude damping is non-unital. Further details are provided in Methods.

\begin{figure}[t]
    \centering
    \includegraphics[width=0.98\linewidth]{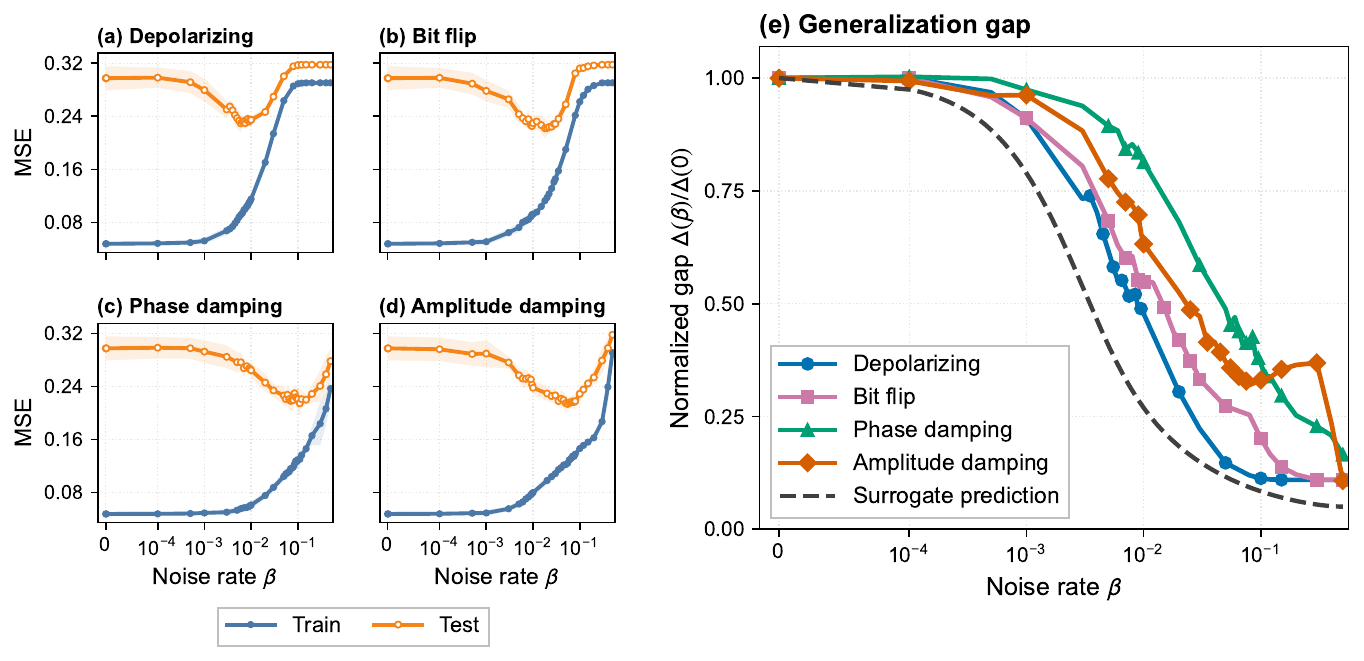}
    \caption{Non-monotonic response of the four-qubit regression model to physical noise. (a--d) Final training and testing mean-squared errors for four quantum noise models. Curves show the mean over ten independently trained runs and shaded bands show one standard deviation. (e) Generalization gap normalized by its value at zero noise. The dashed line shows the surrogate prediction introduced below.}
    \label{fig:nonmonotonic}
\end{figure}

\cref{fig:nonmonotonic} reveals a counterintuitive separation between training and test behavior. Across all four noise channels, the training loss $\mc{L}_{\rm train}(\beta)$ generally increases as the physical noise rate becomes larger. The noisy model therefore fits the training data less accurately. However, the test loss $\mc{L}_{\rm test}(\beta)$ does not follow the same trend. Instead, it first decreases below the nearly noiseless value and only increases again once the noise becomes sufficiently strong. In this finite-noise window, the noisier model performs worse on the data used for training, but better on testing data.

Simple explanations such as optimization quality are not consistent with this behavior. The improvement in test performance is not caused by a lower training loss or by a more successful fit to the training samples. It is instead a change in generalization. We quantify this effect using the generalization gap~\cite{Banchi21_040321,Caro21_582}
\begin{equation}\label{eqn:empirical_gen_gap_main_text}
    \Delta_{\rm gen}(\beta)
    =
    \mc{L}_{\rm test}(\beta)-\mc{L}_{\rm train}(\beta).
\end{equation}
As shown in \cref{fig:nonmonotonic}(e), the gap decreases rapidly in the same intermediate-noise regime where the test loss improves. Thus, finite quantum noise can reduce the discrepancy between training and test performance before the large-noise degradation sets in.

This finite-noise transition is not captured by the two limiting regimes most commonly discussed in QML. In the nearly noiseless regime, the noise-induced effect is absent by construction. In the very noisy regime, noise can strongly suppress useful gradients and degrade trainability. The behavior observed in \cref{fig:nonmonotonic} lies between these limits: noise is already strong enough to change generalization, but not yet strong enough to destroy the learnable signal. This raises a central question: what changes in the QML model when noise is strong enough to improve generalization, but not yet strong enough to destroy the learnable signal? We address this question by introducing a stochastic surrogate model for noisy QML predictions.

\subsection{A noise-order surrogate explains noise-induced regularization}

To understand the performance improvement observed in Fig.~1, we introduce a surrogate model for noisy QML analysis. Consider a circuit with $g$ noisy locations and physical noise rate $\beta$. Each noise realization yields a noise-injected circuit configuration whose response depends on where noise occurs, the noise channel, and the circuit structure. The useful structure is that these many responses can be organized by noise order. Rather than tracking each microscopic configuration separately, we group the noisy circuit responses by the total number of noise occurrences, and represent the combined response at order $m$ by an effective noise-order predictor $\psi_m(\rho_0;\boldsymbol{\theta})$. The overall noisy predictor at physical noise rate $\beta$ is then written as
\begin{equation}
    \phi_{\beta}(\rho_0;\boldsymbol{\theta})
    =
    \sum_{m=0}^{g}
    p_m(\beta,g)\psi_m(\rho_0;\boldsymbol{\theta}),
    \qquad
    p_m(\beta,g)
    =
    \binom{g}{m}\beta^m(1-\beta)^{g-m}.
\end{equation}
This representation has the form of an ensemble average: the predictors $\psi_m$ contain the circuit- and noise-dependent response at each noise order, while the binomial weights $p_m(\beta,g)$ describe how the physical noise rate distributes probability mass across those orders. When $\beta$ is close to zero, most weight is concentrated on the nearly noiseless branch. As $\beta$ increases, the weight spreads over a broader range of finite-noise branches, so the learned predictor is not merely controlled by a single circuit structure.

This surrogate also shows how noise changes the training objective. In the surrogate model, the microscopic noise path is coarse-grained to its occurrence count: an order $m$ is sampled with probability $p_m(\beta,g)$, and the model evaluates the corresponding noise-order predictor $\psi_m$. Let $\mathbb{E}_{\mc{S}}$ denote the empirical average over the training set. As derived in \cref{sec:appendix-A-surrogate-loss}, the surrogate loss can be decomposed as
\begin{equation}\label{eqn:surrogate-loss-bias-var}
    \wt{\mc{L}}(\boldsymbol{\theta},\beta)
    =
    \EE_{\mc{S}}
    \norm{
    \phi_{\beta}(\rho_0;\boldsymbol{\theta})-y
    }^2
    +
    \EE_{\mc{S}}
    \sum_{m_1<m_2}
    p_{m_1}(\beta,g)p_{m_2}(\beta,g)
    \norm{
    \psi_{m_1}(\rho_0;\boldsymbol{\theta})
    -
    \psi_{m_2}(\rho_0;\boldsymbol{\theta})
    }^2 .
\end{equation}
These two terms expose a tradeoff between fitting with the averaged noisy predictor and suppressing variations across noise orders. The first term is the prediction error of the averaged noisy model and reflects the distortion of the model hypothesis space under quantum noise. The second term comes from fluctuations across noisy circuit realizations and penalizes disagreement between predictors at different noise orders. Thus, finite stochastic noise does not only perturb the circuit output; it also favors models whose predictions are more consistent across the noisy orders.

This decomposition gives a mechanism for the separation between training and test behavior in Fig.~1. Noise can worsen the training fit because the averaged noisy predictor is distorted relative to the nearly noiseless circuit. At the same time, the stochastic loss discourages strong sensitivity to the realized noise order, which can improve generalization in an intermediate-noise regime. The next question is how to quantify the strength of this effect as $\beta$ changes. We answer this by introducing the noise-order purity, which measures the concentration of the noise-order distribution and predicts the effective complexity of the noisy model.

\subsection{Noise-order purity predicts effective complexity}

As introduced above, the surrogate model represents the noisy QML outcome as a weighted ensemble of noise-order predictors. This representation separates the circuit-dependent responses $\psi_m$ from the statistical weights $p_m(\beta,g)$. The weights are binomial probabilities determined only by the physical noise rate $\beta$ and the number of noisy locations $g$. As $\beta$ increases from zero, probability mass spreads from the nearly noiseless branch to a range of finite-noise branches. When $g\beta\ll1$, this distribution is well described by a Poisson approximation concentrated near low noise orders, where many-noise occurrences are rare. At larger finite noise, the distribution becomes broader and approaches a Gaussian shape in the central-limit regime. Thus, increasing $\beta$ changes not only the amount of noise, but also the structure of the ensemble being averaged over.

We quantify this spreading by the noise-order purity
\begin{equation}
    s_2(\beta,g)
    =
    \sum_{m=0}^{g}p_m^2(\beta,g).
    \label{eqn:main-s2}
\end{equation}
This quantity is close to one when the noisy predictor is dominated by a single noise order, such as the nearly noiseless branch at small $\beta$. It decreases when the probability mass spreads over many finite-noise orders. Thus, lower noise-order purity corresponds to a stronger averaging effect across noisy circuit responses.

The noise-order purity also gives a quantitative scale for this transition. It is well approximated by
\begin{equation}
    s_2(\beta,g)
    \approx
    e^{-2\beta g} I_0(2\beta g),
    \label{eqn:main-s2-bessel}
\end{equation}
where $I_0$ is the modified Bessel function of the first kind. This expression interpolates between the weak-noise expansion $s_2(\beta,g)=1-2g\beta+\Or(g^2\beta^2)$ for $g\beta\ll1$ and the broader-distribution scaling $s_2(\beta,g)\approx(4\pi g\beta)^{-1/2}$ when $g\beta\gg1$ and $\beta \le 1/2$. Rather than changing at a single rate, $s_2$ shows a regime-dependent decay as the noise-order distribution first leaves the nearly noiseless branch and then spreads over many finite-noise orders. This regime-dependent shape is shown clearly in \cref{fig:nonmonotonic}(e). This crossover provides a quantitative way to describe the finite-noise transition observed in the model performance.

The noise-order purity is also directly connected to the learning-theoretic analysis of the surrogate noisy QML model. As shown by the surrogate-loss decomposition in \cref{eqn:surrogate-loss-bias-var}, stochastic noise induces a regularization term that penalizes disagreement among noise-order predictors. The analysis in \cref{sec:appendix-A-deff} shows that, when different noise-order predictors are weakly correlated, the effective model complexity satisfies
\begin{equation}\label{eqn:deff-main-text}
    d_{\rm eff}(\beta)\propto s_2(\beta,g),
\end{equation}
Although this proportionality is derived under idealized assumptions, it is validated by the numerical results across the QML models studied in this work.

This prediction connects the noise physics to the generalization behavior in Fig.~1. As finite noise spreads probability mass across noise orders, $s_2$ decreases and the effective number of fitted degrees of freedom is reduced. Following the local linearization analysis~\cite{Efron04_619}, the expected generalization gap obeys
\begin{equation}
    \Delta_{\rm gen}(\beta)
    \approx
    \frac{2\sigma_{\rm data}^2}{n_{\rm train}}
    d_{\rm eff}(\beta)
    \propto
    s_2(\beta,g).
    \label{eqn:gen_gap_main_text}
\end{equation}
Here, $\sigma_{\rm data}^2$ is the data-noise variance and $n_{\rm train}$ is the number of training samples. Since these quantities are fixed across the noise sweep, the $\beta$-dependence of the generalization gap is controlled by the effective complexity, and hence by the noise-order purity. This relation links an experimentally measurable quantity, the drop of the generalization gap, to a structural property of the noisy circuit ensemble. Its agreement with the observed decay of $\Delta_{\rm gen}(\beta)$ is shown in \cref{fig:nonmonotonic}(e).

\subsection{Effective-complexity scaling in trained models}

The decreasing generalization gap in \cref{fig:nonmonotonic}(e) provides indirect evidence for the predicted reduction in effective complexity. We test this prediction directly by estimating $d_{\rm eff}$ from the Jacobian of the trained noisy predictor and comparing it with the noise-order purity in \cref{eqn:deff-main-text}.

\begin{figure}[t]
    \centering
    \includegraphics[width=0.95\linewidth]{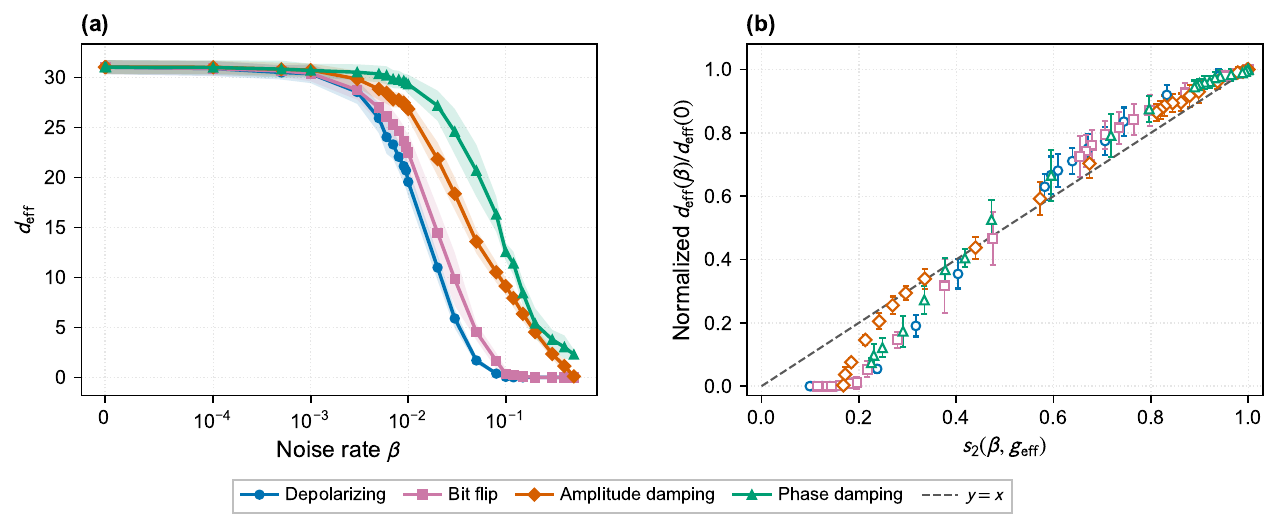}
    \caption{Effective model complexity under physical noise. (a) Effective model complexity as a function of noise rate for the four noise models. Lines show the mean over ten runs and shaded regions show one standard deviation. (b) Effective model complexity normalized by its zero-noise value and plotted against $s_2(\beta,g_{\rm eff})$. Error bars show one standard deviation; fitted values of $g_{\rm eff}$ are reported in \cref{sec:appendix-surrogate-geff}.}
    \label{fig:actual-deff}
\end{figure}

\cref{fig:actual-deff}(a) shows that all four noise models reduce the effective model complexity, but at different rates. We account for this noise-model dependence by fitting $g_{\rm eff}$ separately for each model. After normalization, all four curves follow the predicted dependence on $s_2(\beta,g_{\rm eff})$ in \cref{fig:actual-deff}(b). The ordering of $g_{\rm eff}$ can be understood from how directly each noise model changes the populations probed by the final measurement, which is $Z^{\otimes 4}$ in this case. For example, phase damping suppresses only coherences and therefore affects the measured signal less directly. Depolarizing noise instead changes both populations and coherences, leading to the largest $g_{\rm eff}$. The fitted values and further details are reported in \cref{sec:appendix-surrogate-geff}. This agreement directly validates the predicted scaling $d_{\rm eff}\propto s_2$ and explains the concurrent reduction of the generalization gap in \cref{fig:nonmonotonic}(e).

\subsection{Finite-noise improvement precedes trainability collapse}

The finite-noise improvement identified above occurs in an intermediate regime, and should be distinguished from the very noisy regime where quantum models become difficult to train. Previous studies of noisy QML often emphasize the large-noise limit, where noise suppresses useful gradients and can effectively reduce the trainable circuit depth~\cite{McClean18_4812,Cerezo21_1791,Sharma22_180505,Coles21_6961,Thanasilp23_21,Ragone24_7172,Fontana24_7171,Larocca25_174,Eisert26_751}. This limit is important, but it does not explain the performance improvement observed in \cref{fig:nonmonotonic}, because the improvement occurs before the model enters the gradient-collapse regime.

We evaluate this distinction by computing the parameter-wise output-gradient norm on the training inputs:
\begin{equation}
    \gamma_i(\beta)
    =\sqrt{\mathbb{E}_{\rho_0\sim\mathcal{S}_{\rm train}}
      \left(\partial_{\theta_i}\phi_\beta\!\left(\rho_0;
      \boldsymbol{\theta}^{(\beta)}\right)\right)^2},
\end{equation}
Here, $\boldsymbol{\theta}^{(\beta)}$ denotes the parameters obtained at noise rate $\beta$. Unlike the loss gradient, which also depends on the residual, $\gamma_i$ measures how strongly the model output changes with parameter $i$ and serves as a proxy for trainability.

\begin{figure}[t]
    \centering
    \includegraphics[width=0.98\linewidth]{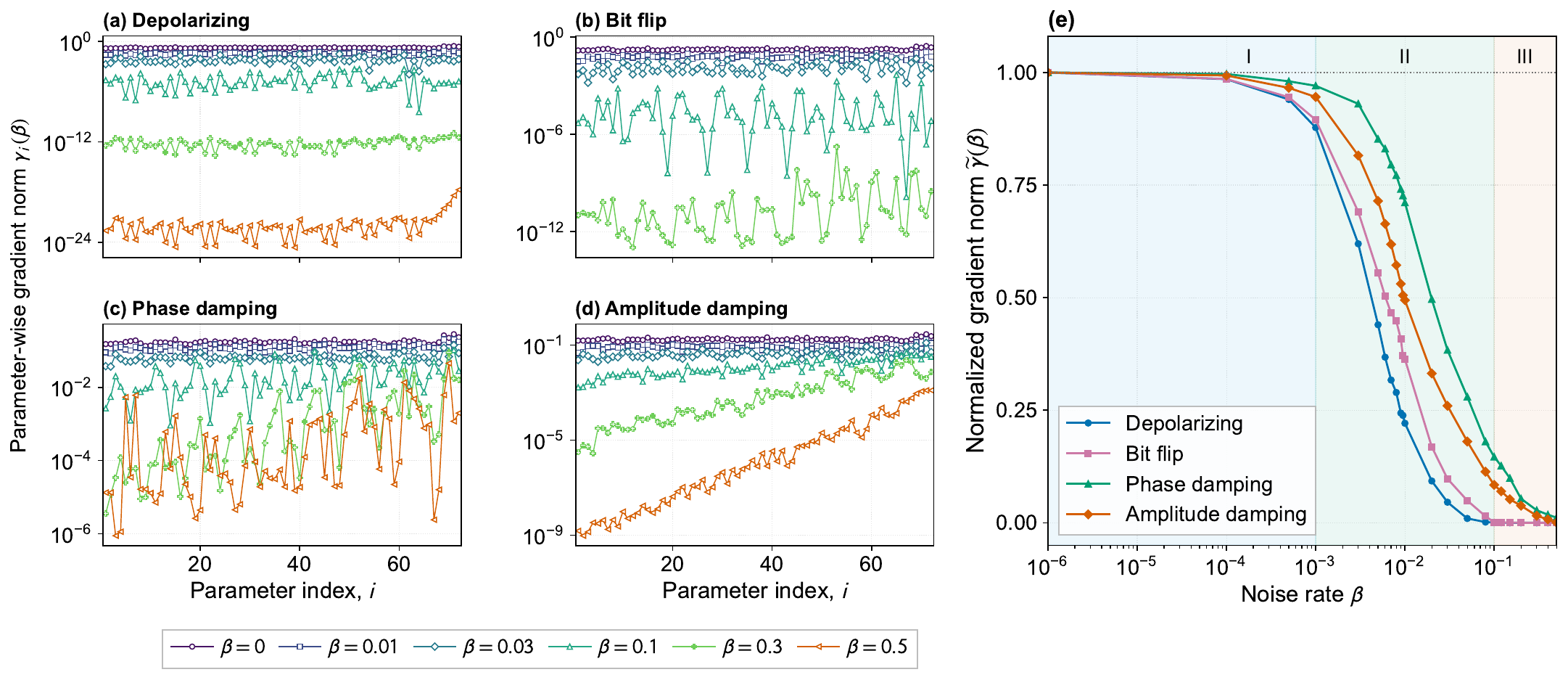}
    \caption{Parameter-wise output-gradient norms under physical noise. (a--d) Root-mean-square output-gradient norm $\gamma_i(\beta)$ versus parameter index $i$ for representative noise rates. Each point is averaged over ten independent runs. In (b), the $\beta=0.5$ bit-flip channel symmetrically averages the identity and bit-flip branches in the final noisy layer, giving $\gamma_i(0.5)=0$ exactly; the corresponding curve is therefore absent on the logarithmic scale. (e) Output-gradient norm averaged over parameters and normalized by its zero-noise value. Roman numerals indicate three noise regimes.}
    \label{fig:gradient-regimes}
\end{figure}

\cref{fig:gradient-regimes}(a--d) shows this quantity for every parameter. For the three unital noise models, the output-gradient norms decrease relatively uniformly across parameters as the noise rate increases. Amplitude damping instead develops a clear dependence on distance from the final measurement: parameters farther from the measurement have smaller output-gradient norms. This trend is consistent with recent results on quantum machine learning under non-unital noise~\cite{Eisert26_751}.

\cref{fig:gradient-regimes}(e) averages $\gamma_i$ over parameters and normalizes it by its zero-noise value. This normalized quantity measures how strongly noise suppresses the output-gradient norm. All four noise models exhibit three regimes. The output-gradient norm changes little in regime I, decreases in regime II, and is strongly suppressed in regime III. Importantly, regime II coincides with the test-error improvement in \cref{fig:nonmonotonic}, while the output-gradient norm remains appreciable.

Together, these results distinguish noise-induced regularization from trainability collapse. In regime II, noise-order averaging reduces effective complexity while the output-gradient norm remains appreciable. Only stronger noise drives the system into regime III, where both training and test errors increase as trainability collapses.

\subsection{Noise programming through controlled parameter perturbations}

\begin{figure}[t]
    \centering
    \includegraphics[width=0.99\linewidth]{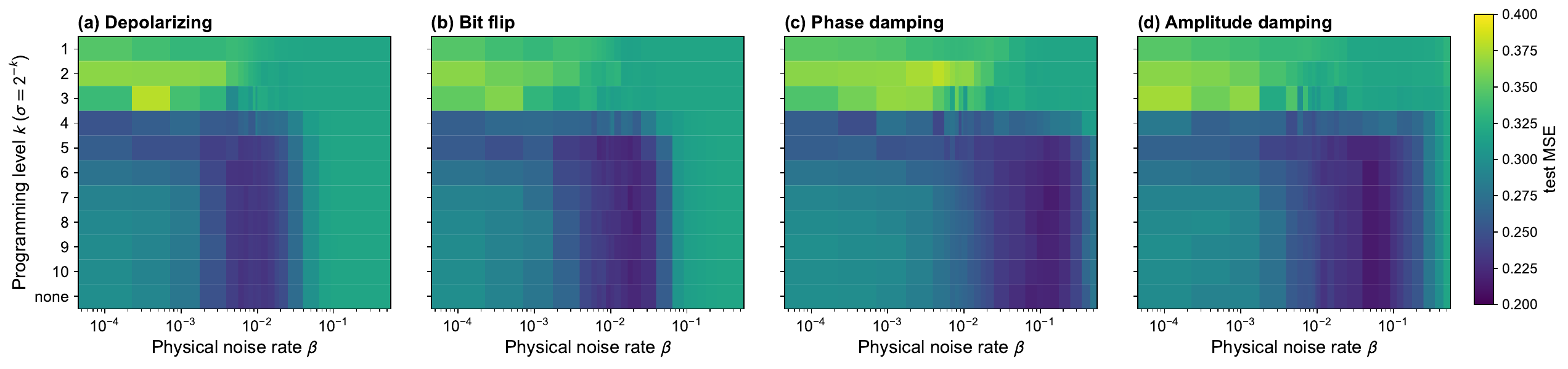}
    \caption{Testing error across physical noise rates and noise-programming levels. Mean testing MSE after training at physical noise rate $\beta$ and programming level $k$, corresponding to a parameter-perturbation strength $\sigma=2^{-k}$. The row labelled ``none'' is the unprogrammed baseline. Each cell is averaged over ten runs.}
    \label{fig:programmability}
\end{figure}

In practice, the native noise level of a quantum device is largely fixed. To exploit the non-monotonic response under this constraint, we introduce a \textit{noise-programming} strategy that injects controlled perturbations into the parameter updates during training. After each optimization step, an independent Gaussian-distributed perturbation with standard deviation $\sigma$ is added to the parameters. This models additional uncertainty in the variational gate parameters. Increasing the programming strength therefore increases the effective noise level experienced by the model. To compare devices at different native noise levels, \cref{fig:programmability} maps the testing error across $\beta$ and $\sigma$.

The horizontal direction recovers the non-monotonic response to native physical noise when the programming perturbation does not dominate. At stronger programming, this structure gradually disappears due to the high effective noise rate. On the other hand, moving along a vertical slice at fixed $\beta$ shows how increasing the programming strength mimics an increase in the effective noise rate for a given device. This intervention is directional: programming can increase the effective noise level but cannot undo the native physical noise. When the unprogrammed model lies on the low-noise side of the optimum, increasing the programming strength can move it towards a higher-noise but lower-error part of the landscape. Once the physical noise has passed the optimum, the same intervention cannot return the model to the lower-noise regime. \cref{fig:programmability} shows this asymmetry for all four physical noise models. This directional accessibility turns the non-monotonic response into a control strategy. Noise programming can therefore move an under-regularized QML model towards the intermediate-noise optimum, where noise-induced regularization improves its predictive performance. Practically, mini-batch stochastic gradient descent generates fluctuations in parameter updates and may therefore provide a natural algorithmic route to noise programming~\cite{Neelakantan15_151106807,Mandt17_1}.

\subsection{Noise-induced regularization persists in molecular QML}

\begin{figure}[t]
    \centering
    \includegraphics[width=0.99\linewidth]{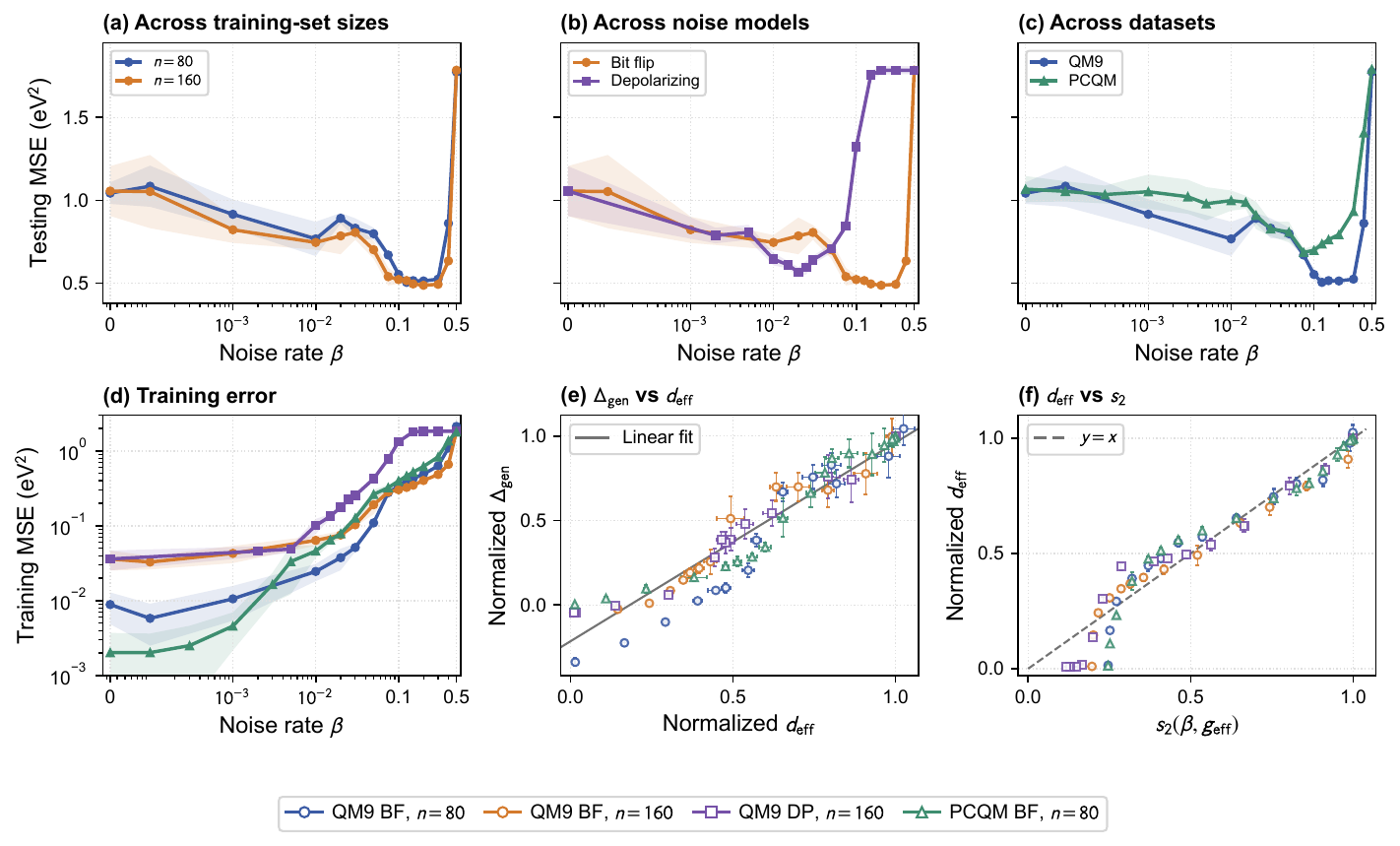}
    \caption{Noise-induced regularization in molecular QML. Testing MSE is compared (a) between $n_{\rm train}=80$ and 160 on QM9-HA8 under bit-flip noise, (b) between bit-flip and depolarizing noise on QM9-HA8 with $n_{\rm train}=160$, and (c) between QM9-HA8 and PCQM4Mv2-HA9 under bit-flip noise with $n_{\rm train}=80$. (d) Training MSE for the four distinct settings in (a--c). (e) Generalization gap versus effective model complexity, with both quantities normalized by their values at $\beta=0$ within each run. (f) Effective model complexity normalized by its value at $\beta=0$ versus $s_2(\beta,g_{\rm eff})$, where $g_{\rm eff}$ is fitted separately for each setting. The QM9-HA8 and PCQM4Mv2-HA9 models are evaluated after 300 and 100 epochs, respectively. Lines and shaded regions in (a--d) show the mean and one standard deviation over five runs; points and error bars in (e,f) show the corresponding mean and standard deviation.}
    \label{fig:molecular-qgnn}
\end{figure}

The four-qubit model allows the finite-noise mechanism to be examined in a setting where the noise response and local model complexity can both be evaluated in detail. We next test whether the same mechanism remains relevant when the predictor contains a larger quantum representation and trainable classical post-processing. Quantum and hybrid quantum--classical models have recently been explored for learning molecular properties and graph representations~\cite{Schuld19_040504,Havlicek19_209,Jerbi23_517,Sajjan22_6475,Rhee23_4300,Lu25_8057}. We study eight- and nine-qubit quantum graph neural networks that predict the HOMO--LUMO gap for molecules drawn from QM9-HA8 and PCQM4Mv2-HA9~\cite{VonLilienfeld14_140022,Hu2021_OGBLSC,Shimazaki17_1300} under gate-level quantum noise. \cref{fig:molecular-qgnn} compares the performance across variable factors including training-set sizes, noise models and molecular datasets.

Across these comparisons in \cref{fig:molecular-qgnn}(a--c), the testing MSE consistently exhibits a finite-noise optimum. The training MSE instead increases with the noise rate (\cref{fig:molecular-qgnn}(d)), indicating the increasing bias introduced by noise. \cref{fig:molecular-qgnn}(e) confirms the linear relation between the generalization gap and the effective model complexity, while \cref{fig:molecular-qgnn}(f) shows that the effective complexity remains proportional to the noise-order purity. These results extend the noise-induced generalization mechanism beyond the four-qubit model: finite noise reduces effective complexity, and the resulting reduction in the generalization gap can improve testing performance before the training error becomes dominant.

\subsection{Finite-noise optima disappear in the large-sample limit}

The finite-noise optimum is not unconditional. Let $\mathrm{SNR}$ denote the strength of the learnable signal relative to the data-noise variance. The analysis in \cref{sec:appendix-A} gives
\begin{equation}\label{eq:finite-noise-optimum}
    s^\star=\frac{\mathrm{SNR}}{\mathrm{SNR}+d_{\rm eff}^{(0)}/n_{\rm train}},\qquad s_2(\beta^\star,g)=s^\star.
\end{equation}
Here, $d_{\rm eff}^{(0)}$ is the effective model complexity in the absence of physical noise. When the data set is given and the SNR is fixed, the optimum is therefore controlled by the ratio $d_{\rm eff}^{(0)} / n_{\rm train}$. Decreasing this ratio moves $s^\star$ towards one and hence moves $\beta^\star$ towards the low-noise limit.

\begin{figure}[t]
    \centering
    \includegraphics[width=0.99\linewidth]{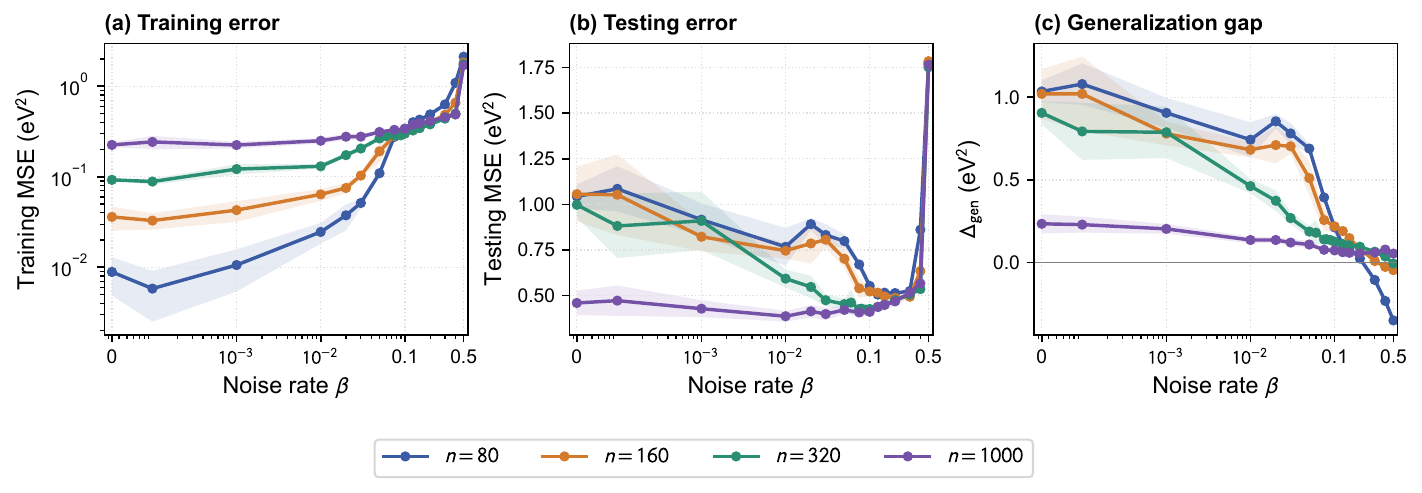}
    \caption{Training-set-size dependence of the finite-noise response. (a) Training MSE, (b) testing MSE and (c) generalization gap under bit-flip noise for nested QM9-HA8 training sets with $n_{\rm train}=80$, 160, 320 and 1000. The first three models are evaluated after 300 epochs and the $n_{\rm train}=1000$ model after 100 epochs. Lines and shaded regions show the mean and one standard deviation over five fixed runs.}
    \label{fig:molecular-training-size}
\end{figure}

We test this shift using nested QM9-HA8 training sets with $n_{\rm train}=80$, 160, 320 and 1000 (\cref{fig:molecular-training-size}). The three smaller training sets retain a clear finite-noise reduction in testing error. For $n_{\rm train}=1000$, the low-noise generalization gap is much smaller and the nonmonotonic response almost disappears. Consistent with \cref{eq:finite-noise-optimum}, the optimum generally shifts towards the low-noise limit as $n_{\rm train}$ increases. The exception is $n_{\rm train}=80$ and 160, for which the optimum occurs at the same sampled noise rate. The model has $q=310$ trainable parameters, so both cases are overparameterized~\cite{Larocca23_542,Peters23_1210,Zhang17_ICLR,Belkin19_15849}. In this regime, the effective model complexity is limited by the lack of training data. Therefore, increasing $n_{\rm train}$ also allows the effective model complexity to grow by exposing additional model directions. Consequently, $d_{\rm eff}^{(0)}/n_{\rm train}$ remains nearly unchanged. Once $n_{\rm train}$ exceeds $q$, this data limitation is removed and the optimum moves towards zero. This is also consistent with the statistical intuition. When the training-set size approaches infinity, the model sees the full data distribution, and therefore the generalization gap vanishes because the testing set contains no additional statistical information. The increasing training error with $n_{\rm train}$ in \cref{fig:molecular-training-size}(a), together with the shrinking generalization gap in \cref{fig:molecular-training-size}(c), reflects the mitigation of overfitting towards the large-sample limit.

\section{Discussion}

In this work, we develop a statistical learning theory for noisy quantum machine learning models. It connects microscopic noise processes to macroscopic learning performance and explains the non-monotonic effect of noise on testing performance. The key quantity is the noise-order purity, which measures how broadly the model response is mixed across noise orders. The intermediate noise regime is very different from the weak- and strong-noise limits. It is not well described by perturbative error accumulation, which controls worst-case deviation, or by trainability collapse, which characterizes the loss of representation. Noise-order purity and the learning-theoretic analysis bridge this missing layer, while the numerical results confirm the separation of the three regimes. The non-monotonic effects reported in recent studies are therefore not occasional artifacts, but a general phenomenon that can be understood and predicted.

Noise is not intrinsically beneficial, and even when it is beneficial, the improvement is not unconditional. Our theoretical analysis shows that the finite-noise optimum behaves differently in over- and underparameterized models and can eventually disappear in the large-sample limit. This analysis combines a noise-order surrogate with a local linearization around the trained optimum. An understanding of noise-induced regularization focusing on optimization dynamics remains open. It is also an open question whether the physical noisy model can be analyzed directly to capture the non-monotonicity without introducing the surrogate.

One intriguing result is that noise programming can move an under-regularized noisy model towards its finite-noise optimum. This opens a new direction for harnessing noise in quantum machine learning based on the learning-theoretic analysis. Different circuit components need not contribute equally to the information used for prediction, and noise acting on them need not have the same effect. Understanding this dependence may help allocate protection towards the components that matter most for prediction in partially fault-tolerant architectures.

The present simulations assume independent gate-level noise. The central hardware question is whether noise-order purity continues to predict model complexity and generalization when errors are correlated or time dependent and circuit outputs are estimated from finite shots. Addressing these effects will extend the prediction and programming of finite-noise optima to more complex and realistic quantum application settings.



\section{Methods}

\subsection{QML models and datasets}

\subsubsection{Four-qubit regression model}

We use the diabetes regression dataset distributed with scikit-learn~\cite{efron2004least,sklearn_diabetes}. From the ten clinical covariates, we retain body mass index and the logarithm of the serum triglyceride level (features 2 and 8 in the dataset). From the 442 samples, we randomly select disjoint training and testing sets of 40 and 400 samples, respectively. The two input features are independently rescaled to $[0,\pi]$, and the regression target is rescaled to $[-1,1]$. Both transformations are fitted on the training set and subsequently applied to the testing set.

We use a four-qubit parametrized quantum circuit adapted from Refs.~\cite{Melnikov25_e00603,Lucchi26_045015}. The circuit comprises nine layers with the same gate layout and independently trainable parameters. In each layer, the two input features are re-uploaded through $R_X$ rotations on qubits 0 and 2. The trainable block then applies $\mathrm{IsingXX}$ rotations to four nearest-neighbour pairs on a periodic ring, followed by one $R_Y$ rotation on each qubit. Each layer therefore contains eight trainable parameters, giving $q=72$ parameters in total. The scalar model output $\phi_\beta(\rho_0;\boldsymbol{\theta})$ is the expectation value of $Z^{\otimes4}$ under the noisy circuit dynamics.

\subsubsection{Molecular QGNN models}

We study molecular regression on fixed-size subsets of QM9 and PCQM4Mv2~\cite{VonLilienfeld14_140022,Hu2021_OGBLSC,Shimazaki17_1300}. QM9-HA8 contains molecules with eight heavy atoms, while PCQM4Mv2-HA9 contains molecules with nine heavy atoms. In both cases, the target is the HOMO--LUMO gap in electronvolts. We construct fixed ordered training pools of 2,000 and 320 molecules for QM9-HA8 and PCQM4Mv2-HA9, respectively, together with disjoint testing sets of 1,500 molecules. All reported training sets are nested prefixes of these pools, and the testing sets are kept fixed across training sizes, noise rates and runs.

Each heavy atom is assigned to one qubit, giving eight- and nine-qubit quantum graph neural networks for QM9-HA8 and PCQM4Mv2-HA9, respectively. Atom features are encoded by single-qubit rotations, and chemical bonds determine bond-type-dependent two-qubit interactions in one EDU-QGC-inspired graph-circuit layer~\cite{Verdon19_190912264,Ceylan21_211205261,Rhee23_4300,Li26_260113275}. We extract the complete computational-basis probability vector from the final quantum state and use a trainable linear readout,
\begin{equation}
    \phi_\beta(x;\boldsymbol{\theta})
    =
    b+\sum_{z}w_z
    \langle z|\rho_\beta(x;\boldsymbol{\theta}_{\rm qc})|z\rangle,
    \qquad
    \boldsymbol{\theta}
    =
    (\boldsymbol{\theta}_{\rm qc},\boldsymbol{w},b).
\end{equation}
Here $x$ denotes the complete molecular input and $\boldsymbol{\theta}_{\rm qc}$ contains the trainable quantum-circuit parameters. The QM9-HA8 model contains 53 quantum-circuit and 257 classical linear-readout parameters, while the PCQM4Mv2-HA9 model contains 56 and 513, respectively, giving 310 and 569 trainable parameters in total. Further circuit and parameter-counting details are given in \cref{sec:appendix-molecular-circuit}.
In \cref{sec:appendix-molecular-readout}, we use a more complex nonlinear classical readout to show that the finite-noise response is not an artifact of the linear readout.

\subsection{Quantum noise models}

We consider four single-qubit noise channels: depolarizing, bit-flip, phase-damping and amplitude-damping noise. At each noisy location, the channel acts on the density matrix through the Kraus map~\cite{Chuang02_Book}
\begin{equation}
    \mathcal{E}_{\beta}(\rho)
    =
    \sum_{\ell}K_{\ell}(\beta)\rho K_{\ell}^{\dagger}(\beta),
\end{equation}
where $\beta\in[0,1]$ is the physical noise rate. In the single-qubit case, the associated Kraus operators are:
\begin{align}
\text{depolarizing:}\quad&
K_0=\sqrt{1-\beta}\,I,\qquad
(K_1,K_2,K_3)=\sqrt{\frac{\beta}{3}}\,(X,Y,Z), \nonumber\\
\text{bit flip:}\quad&
K_0=\sqrt{1-\beta}\,I,\qquad K_1=\sqrt{\beta}\,X, \nonumber\\
\text{phase damping:}\quad&
K_0=
\begin{pmatrix}1&0\\0&\sqrt{1-\beta}\end{pmatrix},
\qquad
K_1=
\begin{pmatrix}0&0\\0&\sqrt{\beta}\end{pmatrix}, \nonumber\\
\text{amplitude damping:}\quad&
K_0=
\begin{pmatrix}1&0\\0&\sqrt{1-\beta}\end{pmatrix},
\qquad
K_1=
\begin{pmatrix}0&\sqrt{\beta}\\0&0\end{pmatrix}.
\label{eq:methods-kraus}
\end{align}

In all models, the selected channel is applied after every elementary gate in the data encoding and trainable circuit. After a two-qubit gate, the single-qubit channel is applied independently to both participating qubits. No noise is applied to the classical post-processing including the regression. Noiseless circuits are evaluated with state-vector simulation, while noisy circuits are evaluated with the density-matrix simulator in PennyLane~\cite{bergholm2018pennylane}. The four-qubit circuit has a fixed number of noisy locations, whereas the molecular circuit follows the input graph and therefore has a molecule-dependent count. The corresponding gate accounting is given in \cref{sec:appendix-molecular-noise-count}.

\subsection{Training and evaluation}

The four-qubit model is trained by minimizing the mean-squared error between its scalar output and the regression target. We use full-batch Adam optimization~\cite{kingma2014adam} with learning rate $0.01$ for 800 epochs and do not use early stopping. The physical noise channel at the selected value of $\beta$ is included during both training and evaluation. Testing samples are not used to update the parameters.

For each physical noise model and noise rate, we train ten independently initialized models. The 72 initial parameters in each run are sampled from a standard normal distribution using a fixed random seed. The same set of ten initial parameter vectors is reused across all noise rates and noise models, so changes across the noise sweep are not caused by different initializations. Training and testing metrics are evaluated after the final update. Reported curves show the mean over the ten runs, and uncertainty bands or error bars show one sample standard deviation across runs.

The molecular models are trained end to end by minimizing the mean-squared error in electronvolts squared. We use Adam with learning rate $0.03$, a mini-batch size of 16 and gradient clipping with maximum norm 1. Each epoch visits every training molecule once in a deterministically shuffled order. The readout bias is initialized to the mean target in the training set. QM9-HA8 and PCQM4Mv2-HA9 models are trained for 300 and 100 epochs, respectively, and all metrics are evaluated at the final checkpoint.

For each molecular setting, five fixed initializations are used across the complete noise sweep. The physical noise channel is included during both training and evaluation, and every final checkpoint is evaluated on all 1,500 testing molecules. Molecular curves show the mean over the five runs, with uncertainty given by one sample standard deviation.

\subsection{Local model complexity and effective noise count}

We estimate model complexity by linearizing the physical noisy predictor around the trained parameters. For run $r$,
\begin{equation}
    \phi_\beta(\rho_a;\boldsymbol{\theta}^{(\beta,r)}+\delta\boldsymbol{\theta})
    \approx
    \phi_\beta(\rho_a;\boldsymbol{\theta}^{(\beta,r)})
    +
    \sum_{i=1}^{q}J_{ai}^{(r)}(\beta)\delta\theta_i ,
    \qquad
    J_{ai}^{(r)}(\beta)
    =
    \left.
    \frac{\partial\phi_\beta(\rho_a;\boldsymbol{\theta})}
         {\partial\theta_i}
    \right|_{\boldsymbol{\theta}=\boldsymbol{\theta}^{(\beta,r)}} .
\end{equation}
The nonlinear model therefore reduces locally to a linear regression problem with the predictor Jacobian $J^{(r)}(\beta)$ as its design matrix. Let $\lambda_j^{(r)}(\beta)$ denote the eigenvalues of $J^{(r)}(\beta)^\mathsf{T}J^{(r)}(\beta)/n_{\rm train}$. These eigenvalues quantify how strongly the model output responds along different parameter directions. A hard count of nonzero eigenvalues is sensitive to numerical precision and treats directions of very different strengths equally. We instead use the soft count~\cite{Ye98_120,Efron04_619}
\begin{equation}
    d_{\rm eff}^{(r)}(\beta)
    =
    \sum_{j=1}^{q}
    \frac{\lambda_j^{(r)}(\beta)}
         {\lambda_j^{(r)}(\beta)+10^{-4}}
    \approx
    \sum_{j=1}^{q}
    \mathbb{I}\!\left(\lambda_j^{(r)}(\beta)\neq0\right) .
\label{eq:methods-deff}
\end{equation}
Directions with eigenvalues well above $10^{-4}$ contribute approximately one, whereas weaker directions contribute proportionally less. The same counting scale is used for every noise model, noise rate and run.

For the molecular models, $J^{(r)}(\beta)$ is taken with respect to the complete parameter vector $\boldsymbol{\theta}$, including both the quantum-circuit and linear-readout parameters.

For comparisons across noise models, each run is normalized by its own zero-noise value, $\widetilde d_{\rm eff}^{(r)}(\beta)=d_{\rm eff}^{(r)}(\beta)/d_{\rm eff}^{(r)}(0)$, before averaging across runs. We obtain the effective noise count by fitting this averaged curve to the exact binomial noise-order purity,
\begin{equation}
    g_{\rm eff}
    =
    \underset{g\in\{1,\ldots,1000\}}{\operatorname{argmin}}\,
    \frac{1}{N_\beta}
    \sum_{\beta}
    \left(
      s_2(\beta,g)
      -
      \frac{1}{R}\sum_{r=1}^{R}
      \widetilde d_{\rm eff}^{(r)}(\beta)
    \right)^2 ,
\end{equation}
where $s_2(\beta,g)$ is the noise-order purity defined in \cref{eqn:main-s2}, and $R=10$ and $5$ for the four-qubit and molecular experiments, respectively. The dashed reference curve in \cref{fig:nonmonotonic}(e) uses the exact four-qubit noisy-location count $g=126$, rather than a fitted count. The fitted $g_{\rm eff}$ summarizes how strongly a noise model changes the measured predictor across these locations. Its channel dependence and order-resolved numerical validation are discussed in \cref{sec:appendix-surrogate}.

The parameter-wise output-gradient norm used in \cref{fig:gradient-regimes} is computed from the same predictor Jacobian:
\begin{equation}
    \gamma_i^{(r)}(\beta)
    =
    \sqrt{
      \frac{1}{n_{\rm train}}
      \sum_{a=1}^{n_{\rm train}}
      \left(J_{ai}^{(r)}(\beta)\right)^2
    }.
\end{equation}
The profiles in \cref{fig:gradient-regimes}(a--d) show $\sum_{r=1}^{10} \gamma_i^{(r)}(\beta) / 10$. For the scalar response in \cref{fig:gradient-regimes}(e), we average $\gamma_i^{(r)}$ over both parameters and runs and divide by the corresponding average at $\beta=0$.

\subsection{Noise programming}

Noise programming is implemented by adding an independent Gaussian perturbation after every Adam update. If $\Delta_t^{\rm Adam}$ denotes the optimizer update at step $t$, the programmed parameters obey
\begin{equation}
    \boldsymbol{\theta}_{t+1}
    =
    \boldsymbol{\theta}_{t}
    +\Delta_t^{\rm Adam}
    +\sigma\boldsymbol{\xi}_{t},
    \qquad
    \boldsymbol{\xi}_{t}\sim\mathcal{N}(0,I_q).
\label{eq:methods-programming}
\end{equation}
A new perturbation is sampled independently at every update and for every run. We use programming strengths $\sigma=2^{-k}$ for $k=1,\ldots,10$, together with an unprogrammed baseline $\sigma=0$. All other training settings, including the initial parameters and the physical noise channel, are held fixed.

For every physical noise rate $\beta$ and programming level $k$, the model is retrained for 800 full-batch epochs in each of the ten runs. Each cell in the programming landscape is the final testing MSE averaged across these runs.

\begingroup
\makeatletter
\let\addcontentsline\@gobblethree
\makeatother

\section*{Data availability}

The diabetes dataset is distributed with scikit-learn~\cite{efron2004least,sklearn_diabetes}. QM9~\cite{VonLilienfeld14_140022} and PCQM4Mv2~\cite{Hu2021_OGBLSC,Shimazaki17_1300} are publicly available from their original sources. The fixed molecular subsets and data splits used in this study, the processed numerical data underlying the figures and the trained checkpoints are available at \url{https://github.com/dongsnaq/Finite-Noise-Generalization-QML}.

\section*{Code availability}

The code developed in this work is available via GitHub at \url{https://github.com/dongsnaq/Finite-Noise-Generalization-QML}.

\section*{Acknowledgements}

The machine learning research was supported by the Office of Naval Research under MURI Award No. N00014-23-1-2001. The quantum computing research was supported by the U.S. National Science Foundation through the Chemical Theory, Models, and Computational Methods (CTMC) program under Grant No. CHE-2515209 to X.L.
This research used resources of the National Energy Research Scientific Computing Center, a DOE Office of Science User Facility supported by the Office of Science of the U.S. Department of Energy under Contract No. DE-AC02-05CH11231, using NERSC awards DDR-ERCAP0038972 and DDR-ERCAP0038957 (Y.D.).

\section*{Author contributions}

Y.D. and X.L. conceived the project. Y.D. developed the theory, designed and performed the numerical study, analyzed the results and prepared the original manuscript. Z.J. developed the initial four-qubit simulation, which Z.Z. subsequently extended. Z.Z. also developed the molecular QML simulation framework. Both Z.Z. and Z.J. contributed to the numerical study and initial explorations. All authors contributed to the discussion and writing of the manuscript.

\section*{Competing interests}

The authors declare no competing interests.

\endgroup

\clearpage
\newpage
\appendix
\renewcommand{\thesubsection}{\thesection.\arabic{subsection}}

\tableofcontents

\section{Theory of noise-induced regularization}
\label{sec:appendix-A}
\subsection{Problem setup and stochastic noise-occurrence expansion}
\label{sec:appendix-A-setup}

In this section, we introduce the setup of the problems we consider in this work. By introducing a stochastic noise-occurrence expansion, we discuss a representation of QML models in the presence of noise and use it as a surrogate model to understand the mechanism behind non-monotonic noise response.

The QML model is implemented in terms of a sequence of parametrized gates $\{U_1(\theta_1), \cdots, U_g(\theta_g)\}$. The input density matrix $\rho_0$ carries the information of the data point, such as encodings of chemical structures and molecular features. The idealized output of such a parametrized circuit is
\begin{equation}
    \rho_{\rm out}^{(0)} = \mc{U}_g \circ \cdots \circ \mc{U}_1(\rho_0),
\end{equation}
where $\mc{U}_j(\rho) = U_j(\theta_j) \rho U_j^\dagger(\theta_j)$ is the corresponding conjugate action. For simplicity, we assume that the data enter through the initial state $\rho_0$. More general data-dependent gate maps leave the analysis unchanged and are reflected in the model-dependent quantities appearing in the final results. The final quantum state is measured with some physical observable, for example an observable $O$, so that the information is extracted as
\begin{equation}
    \phi_0(\rho_0; \boldsymbol{\theta}) := \Tr(O\rho_{\rm out}^{(0)}).
\end{equation}

In the presence of noise, we assume that after each ideal gate action there is an additional quantum channel distorting the idealized dynamics. Suppose the noise channels are $\mc{E}_1, \cdots, \mc{E}_g$. The noisy output is
\begin{equation}
    \rho_{\rm out}^{\mc{E}} = \mc{E}_g \circ \mc{U}_g \circ \cdots \circ \mc{E}_1 \circ \mc{U}_1(\rho_0),
\end{equation}
and the corresponding noisy predictor is
\begin{equation}
    \phi_{\mc{E}}(\rho_0; \boldsymbol{\theta}) := \Tr(O\rho_{\rm out}^{\mc{E}}).
\end{equation}
We consider a training set $\mc{S}=\{(\rho_i,y_i)\}_{i=1}^n$ and the corresponding mean square loss:
\begin{equation}
    \mc{L}(\boldsymbol{\theta}) 
    = \EE_{\mc{S}}\norm{\phi_{\mc{E}}(\rho_0; \boldsymbol{\theta}) - y}^2.
\end{equation}

To understand the non-monotonic noise response, we introduce a stochastic noise-occurrence model to study the mechanism. We first discuss the simplest model to show the intuition. Nevertheless, the mechanism does not only apply to this simple model. We discuss the general form of the model and the corresponding extension at the end.

Consider the simplest case where the noise channel at each location either does nothing or applies another noisy operation. This captures, for example, dephasing or bit-flip noise, where an additional Pauli operator is applied at random. In this case, each noise realization generates an effective circuit that deviates from the idealized circuit by one or more noise occurrences. Therefore, the noise realization can be described by a stochastic process, and the corresponding predictor is associated with a particular noise path. This motivates a stochastic surrogate model in which disagreement among noise-induced predictors appears as an explicit variance term. Such a representation provides a mechanism by which moderate noise can reduce effective model complexity and improve generalization, while excessive noise can destroy learnable signal.

To quantify this intuition, we consider a stochastic noise model where the noise at the $j$-th location is associated with a Bernoulli random variable $z_j \sim \mathrm{Ber}(\beta)$. Conditioned on $z_j$, the inserted channel is
\begin{equation}\label{eqn:stochastic-noise-model}
    \mc{E}_{j,z_j} = (1 - z_j) \mathrm{Id} + z_j \mc{N}_j =
    \begin{cases}
    \mathrm{Id}, & z_j = 0,\\
    \mc{N}_j, & z_j = 1,
    \end{cases}
\end{equation}
where $\mc{N}_j$ is another noise channel, such as a conjugate action of an additional gate. The averaged single-location channel is therefore
\begin{equation}
    \EE_{z_j}\mc{E}_{j,z_j} = (1-\beta)\mathrm{Id} + \beta \mc{N}_j.
\end{equation}
For a noise realization $\boldsymbol{z} = (z_1,\cdots,z_g)$, the conditional noisy predictor is
\begin{equation}
    \phi_{\boldsymbol{z}}(\rho_0; \boldsymbol{\theta}) 
    :=
    \Tr\left(
    O\,
    \mc{E}_{g,z_g} \circ \mc{U}_g \circ \cdots \circ
    \mc{E}_{1,z_1} \circ \mc{U}_1(\rho_0)
    \right).
\end{equation}
Let
\begin{equation}
    c(\boldsymbol{z}) = \sum_{j=1}^g z_j
\end{equation}
denote the number of noise occurrences. We define the occurrence-$m$ predictor as the average over all noise realizations with exactly $m$ noise occurrences:
\begin{equation}
    \psi_m(\rho_0; \boldsymbol{\theta})
    :=
    \EE_{\boldsymbol{z}}
    \left(
    \phi_{\boldsymbol{z}}(\rho_0; \boldsymbol{\theta})
    \mid
    c(\boldsymbol{z}) = m
    \right).
\end{equation}
The noisy mean predictor induced by this stochastic noise-occurrence model can then be expanded with respect to the occurrence count:
\begin{equation}
\begin{split}
    \phi_\beta(\rho_0; \boldsymbol{\theta})
    &:= \EE_{\boldsymbol{z}}
    \left(
    \phi_{\boldsymbol{z}}(\rho_0; \boldsymbol{\theta})
    \right) \\
    &= \sum_{m=0}^g 
    \mathbb{P}(c(\boldsymbol{z}) = m)
    \psi_m(\rho_0; \boldsymbol{\theta}) \\
    &= \sum_{m=0}^g 
    \binom{g}{m}\beta^m(1-\beta)^{g-m}
    \psi_m(\rho_0; \boldsymbol{\theta}).
\end{split}
\end{equation}
Here, the formalism can be viewed as a noise-occurrence expansion: the basis elements are effective predictors generated by different noise occurrence counts, and the expansion weights are determined by the binomial distribution of the occurrence count.

We consider the following stochastic loss as a surrogate model for analysis:
\begin{equation}
    \wt{\mc{L}}(\boldsymbol{\theta}, \beta) 
    = \EE_{\mc{S}, \boldsymbol{z}}
    \norm{
    \phi_{\boldsymbol{z}}(\rho_0; \boldsymbol{\theta}) - y
    }^2.
\end{equation}
This surrogate loss is useful because it exposes an explicit variance term across noise realizations. By the bias-variance identity, we have
\begin{equation}\label{eqn:qml-bias-var-decomp}
\begin{split}
    \wt{\mc{L}}(\boldsymbol{\theta}, \beta)
    &= \EE_{\mc{S}, \boldsymbol{z}}
    \norm{
    \phi_{\boldsymbol{z}}(\rho_0; \boldsymbol{\theta}) - y
    }^2 \\
    &= \EE_{\mc{S}}
    \norm{
    \phi_\beta(\rho_0; \boldsymbol{\theta}) - y
    }^2
    +
    \EE_{\mc{S}, \boldsymbol{z}}
    \mathrm{Var}_{\boldsymbol{z}}
    \left(
    \phi_{\boldsymbol{z}}(\rho_0; \boldsymbol{\theta})
    \right).
\end{split}
\end{equation}
The first term is the mean-channel loss associated with the averaged noisy predictor $\phi_\beta$, while the second term measures the disagreement among predictors generated by different noise realizations. Thus, in the stochastic surrogate model, disagreement among noise realizations appears as an explicit variance penalty. In the noisy mean predictor itself, the same noise-occurrence structure acts through the ensemble average $\phi_\beta = \sum_m p_m \psi_m$, which filters the accessible function class and provides a mechanism for noise-induced regularization.

\subsection{Noise-induced regularization in the surrogate loss}
\label{sec:appendix-A-surrogate-loss}

In this subsection, we analyze the regularization term in \cref{eqn:qml-bias-var-decomp}. Recall that
\begin{equation}
    p_m(\beta,g) := \mathbb{P}(c(\boldsymbol{z}) = m)
    = \binom{g}{m}\beta^m(1-\beta)^{g-m},
\end{equation}
where $c(\boldsymbol{z})=\sum_{j=1}^g z_j$ is the number of noise occurrences. The variance term in the stochastic surrogate loss is
\begin{equation}
    \begin{split}
        \mc{R}(\boldsymbol{\theta}; \beta, g)
        :=& \EE_{\mc{S}}\mathrm{Var}_{\boldsymbol{z}}
        \left(\phi_{\boldsymbol{z}}(\rho_0; \boldsymbol{\theta})\right) = \EE_{\mc{S}, \boldsymbol{z}}
        \norm{\phi_{\boldsymbol{z}}(\rho_0; \boldsymbol{\theta})
        - \phi_\beta(\rho_0; \boldsymbol{\theta})}^2 \\
        =& \EE_{\mc{S}, \boldsymbol{z}}
        \norm{\sum_{m=0}^g
        \left(\mathbb{I}_{c(\boldsymbol{z}) = m}
        - p_m(\beta,g)\right)
        \psi_m(\rho_0; \boldsymbol{\theta})}^2 \\
        =& \EE_{\mc{S}}
        \sum_{m_1,m_2=0}^g
        \psi_{m_1}(\rho_0; \boldsymbol{\theta})
        \psi_{m_2}(\rho_0; \boldsymbol{\theta})
        \Lambda_{m_1,m_2}(\beta,g).
    \end{split}
\end{equation}
Here, the covariance matrix $\Lambda(\beta,g)$ is defined by
\begin{equation}
    \Lambda_{m_1,m_2}(\beta,g)
    := \EE_{\boldsymbol{z}}
    \left(\mathbb{I}_{c(\boldsymbol{z}) = m_1}
    \mathbb{I}_{c(\boldsymbol{z}) = m_2}\right)
    - p_{m_1}(\beta,g)p_{m_2}(\beta,g)
    = \delta_{m_1,m_2}p_{m_1}(\beta,g)
    - p_{m_1}(\beta,g)p_{m_2}(\beta,g),
\end{equation}
where $\delta_{m_1,m_2}$ is the Kronecker delta symbol. 
It can be shown that the covariance matrix $\Lambda(\beta,g)$ is positive semidefinite. 
The regularization term can also be written as a pairwise disagreement penalty:
\begin{equation}
    \mc{R}(\boldsymbol{\theta}; \beta, g)
    =
    \frac{1}{2}
    \EE_{\mc{S}}
    \sum_{m_1,m_2=0}^g
    p_{m_1}(\beta,g)p_{m_2}(\beta,g)
    \norm{\psi_{m_1}(\rho_0; \boldsymbol{\theta})
    - \psi_{m_2}(\rho_0; \boldsymbol{\theta})}^2.
\end{equation}
This form shows that the stochastic surrogate loss penalizes disagreement among predictors associated with different noise occurrence counts.

The trace of the covariance matrix is
\begin{equation}
    \Tr(\Lambda(\beta,g))
    =
    1-\sum_{m=0}^g p_m^2(\beta,g)
    =:
    1-s_2(\beta,g),
\end{equation}
where the noise-order purity
\begin{equation}
\label{eqn:s_2}
    s_2(\beta,g)
    :=
    \sum_{m=0}^g p_m^2(\beta,g)
\end{equation}
measures the concentration of the noise occurrence-count distribution. Therefore, $1-s_2(\beta,g)$ measures the total variance of this distribution. When the occurrence-count distribution spreads over more values of $m$, $s_2(\beta,g)$ decreases and the total variance captured by $\Lambda(\beta,g)$ increases. In the next subsection, we show that the same concentration quantity $s_2(\beta,g)$ also controls the local sensitivity of the averaged noisy predictor.

\subsection{Noise-induced generalization effect and reduced effective model complexity}
\label{sec:appendix-A-deff}

In this subsection, we analyze how the noise-induced regularization term affects the generalization of the stochastic surrogate model. We work with the occurrence-count coarsened surrogate, where a noise realization with $c(\boldsymbol{z})=m$ is represented by the predictor $\psi_m$. This is the surrogate model used to expose the effect of noise-occurrence averaging on model complexity.

In this subsection, we slightly overload notation and write $\psi_m(\boldsymbol{\theta})\in\mathbb{R}^n$ for the vector of predictions $(\psi_m(\rho_i;\boldsymbol{\theta}))_{i=1}^n$, and $y\in\mathbb{R}^n$ for the vector of labels. The averaged noisy predictor on the training set is
\begin{equation}
    \phi(\boldsymbol{\theta},\beta)=\sum_{m=0}^g p_m(\beta,g)\psi_m(\boldsymbol{\theta}).
\end{equation}
Using the decomposition in \cref{eqn:qml-bias-var-decomp}, the stochastic surrogate loss can be written as
\begin{equation}
    \wt{\mc{L}}(\boldsymbol{\theta},\beta)=\frac{1}{n}\norm{\phi(\boldsymbol{\theta},\beta)-y}^2+\frac{1}{n}\sum_{m=0}^g p_m(\beta,g)\norm{\psi_m(\boldsymbol{\theta})}^2-\frac{1}{n}\norm{\phi(\boldsymbol{\theta},\beta)}^2.
\end{equation}
Here, division by $n$ makes the squared vector norms equivalent to empirical averages over $\mc{S}$.
Suppose there are $q$ trainable parameters. The parameter count $q$ and the number of noisy locations $g$ need not coincide. For example, in the four-qubit model studied here, $q=72$ and $g=126$. Let
\begin{equation}\label{eqn:order-jacobians}
J_m=\nabla_{\boldsymbol{\theta}}\psi_m(\boldsymbol{\theta})\in\mathbb{R}^{n\times q}, \qquad J=\nabla_{\boldsymbol{\theta}}\phi(\boldsymbol{\theta},\beta)=\sum_{m=0}^g p_m(\beta,g)J_m.
\end{equation}
The first-order condition (FOC) of the surrogate loss is
\begin{equation}
\nabla_{\boldsymbol{\theta}}\wt{\mc{L}}(\boldsymbol{\theta},\beta)=\frac{2}{n}\sum_{m=0}^g p_m(\beta,g)J_m^\top(\psi_m-y)=0.
\end{equation}
This is equivalent to differentiating the decomposed form, since
\begin{equation}
    J^\top(\phi-y)+\sum_{m=0}^g p_m(\beta,g)J_m^\top\psi_m-J^\top\phi=\sum_{m=0}^g p_m(\beta,g)J_m^\top(\psi_m-y).
\end{equation}
Now we differentiate the FOC with respect to the label vector $y$. Under the local linear approximation, we assume that the Jacobians $J_m$ are stable in a neighborhood of the optimum and ignore the second-order residual terms. Then
\begin{equation}
    \sum_{m=0}^g p_m(\beta,g)J_m^\top(J_m\ud\boldsymbol{\theta}-\ud y)=0.
\end{equation}
Therefore,
\begin{equation}
    \ud\boldsymbol{\theta}=\left(\sum_{m=0}^g p_m(\beta,g)J_m^\top J_m\right)^{-1}J^\top\ud y.
\end{equation}
Since $\ud\phi=J\ud\boldsymbol{\theta}$, the local response of the fitted predictor to label perturbations is
\begin{equation}\label{eqn:local-hat-matrix}
    \ud\phi=J\left(\sum_{m=0}^g p_m(\beta,g)J_m^\top J_m\right)^{-1}J^\top\ud y=:H(\boldsymbol{\theta};\beta,g)\ud y.
\end{equation}
The matrix $H(\boldsymbol{\theta};\beta,g)$ is the local hat matrix of the stochastic surrogate model. Unlike the ordinary least-squares hat matrix, it is generally not an exact projector; its trace plays the role of an effective number of fitted degrees of freedom.

Under the standard regression model with homoscedastic data noise $y=f+\epsilon$, $\EE(\epsilon)=0$, and $\mathrm{Cov}(\epsilon)=\sigma^2 I_n$, the expected generalization gap of a local linear smoother~\cite{Efron04_619} is approximated by
\begin{equation}\label{eqn:generalization_gap}
    \Delta_{\rm gen}:=\EE(\mc{L}_{\rm test}-\mc{L}_{\rm train})\approx\frac{2\sigma^2}{n}\Tr(H)=:\frac{2\sigma^2}{n}d_{\rm eff}.
\end{equation}
Here, $d_{\rm eff}:=\Tr(H)$ is the effective model complexity of the stochastic surrogate model.

To connect this effective complexity to the noise occurrence distribution, we consider a simplified overlap model for the Jacobians. Suppose that the occurrence-count predictors have comparable local sensitivities and approximately uniform cross-overlap:
\begin{equation}
    J_m^\top J_m\approx cI_q,\qquad J_{m_1}^\top J_{m_2}\approx c\kappa I_q \quad (m_1\neq m_2),
\end{equation}
where $c>0$ controls the overall local sensitivity and $\kappa\in[0,1]$ measures the average alignment between different occurrence-count predictors. Then
\begin{equation}
    \sum_{m=0}^g p_m J_m^\top J_m\approx cI_q.
\end{equation}
Moreover,
\begin{equation}
    J^\top J=\sum_{m_1,m_2=0}^g p_{m_1}p_{m_2}J_{m_1}^\top J_{m_2}\approx c\left(s_2(\beta,g)+\kappa(1-s_2(\beta,g))\right)I_q.
\end{equation}
Therefore,
\begin{equation}
    d_{\rm eff}=\Tr(H)\approx q\left(s_2(\beta,g)+\kappa(1-s_2(\beta,g))\right)=q\left(\kappa+(1-\kappa)s_2(\beta,g)\right).
\end{equation}
When different occurrence-count predictors are weakly aligned, $\kappa\approx0$, this reduces to
\begin{equation}\label{eqn:d_eff_s2}
    d_{\rm eff}\approx q s_2(\beta,g).
\end{equation}
Thus, when the noise occurrence-count distribution spreads over more values of $m$, $s_2(\beta,g)$ decreases and the stochastic surrogate model has smaller effective complexity. For a QML model with $q$ trainable parameters, the above analysis shows that the effective model complexity of the stochastic surrogate model is reduced by a factor $s_2(\beta,g)$. This provides a mechanism by which moderate noise can reduce the generalization gap. At the same time, stronger noise generally increases the training loss by attenuating the learnable signal and increasing the approximation bias. The competition between the decreasing generalization gap and the increasing training loss naturally leads to a non-monotonic testing loss. We will provide a more quantitative analysis in the next subsection.

\subsection{Scaling regimes of the noise-order purity}

According to \cref{eqn:d_eff_s2}, the effective model complexity is approximately reduced by the factor $s_2(\beta,g)$. We refer to $s_2(\beta,g)$ as the noise-order purity:
\begin{equation}
    s_2(\beta,g):=\sum_{m=0}^g p_m^2(\beta,g).
\end{equation}
When $\beta=0$, the only possible noise order is the noiseless one, and hence $s_2(\beta,g)=1$. As the noise strength increases, the probability distribution spreads over more noise orders, and $s_2(\beta,g)$ decreases. In our simplified model, the noise order follows a binomial distribution, which allows a quantitative analysis of $s_2(\beta,g)$.

We first analyze two limiting regimes. In the weak-noise regime $g\beta\ll 1$, the noise-order distribution is concentrated near zero. Therefore,
\begin{equation}
    s_2(\beta,g)=\sum_{m=0}^g p_m^2(\beta,g)=p_0^2(\beta,g)+\Or(g^2\beta^2)=1-2g\beta+\Or(g^2\beta^2).
\end{equation}
This shows that very weak noise only induces a perturbative reduction in the effective model complexity.

In the moderate-noise regime, where $g\beta\gg 1$, the binomial distribution can be approximated by a Gaussian distribution with mean $\mu=g\beta$ and variance $\sigma^2=g\beta(1-\beta)$. Hence,
\begin{equation}
    s_2(\beta,g)\approx \frac{1}{2\pi\sigma^2}\int_{-\infty}^{\infty}e^{-(x-\mu)^2/\sigma^2}\ud x=\frac{1}{2\sqrt{\pi}\sigma}=\frac{1}{2\sqrt{\pi g\beta(1-\beta)}}.
\end{equation}
Thus, when the noise-order distribution is sufficiently spread out, the noise-order purity decreases as $1/\sqrt{g\beta(1-\beta)}$.

The transition between these regimes can also be captured by a Poisson approximation with rate $g \beta$. Therefore,
\begin{equation}
    s_2(\beta,g)\approx e^{-2g\beta}\sum_{m=0}^{\infty}\frac{(g\beta)^{2m}}{(m!)^2}=e^{-2g\beta}I_0(2g\beta),
\end{equation}
where $I_0$ is the modified Bessel function of the first kind. This expression interpolates between the weak-noise expansion $s_2(\beta,g)=1-2g\beta+\Or(g^2\beta^2)$ when $g\beta\ll1$, and the Gaussian scaling $s_2(\beta,g)\approx(4\pi g\beta)^{-1/2}$ when $g\beta\gg1$ and $\beta\ll1$.

These scaling regimes have a direct implication for the effective complexity of the stochastic surrogate model. For simplicity, assume that the number of trainable parameters is $q=g$. In the weak-noise regime, $s_2(\beta,g)\approx1$, so the effective complexity remains $d_{\rm eff}\approx g$. In contrast, in the moderate-noise regime,
\begin{equation}
    d_{\rm eff}\approx g s_2(\beta,g)\approx \frac{\sqrt{g}}{2\sqrt{\pi\beta(1-\beta)}}.
\end{equation}
Thus, noise can reduce the effective model complexity from $\Or(g)$ to $\Or(\sqrt{g})$ when the noise-order distribution is sufficiently spread out. Since the generalization gap of the stochastic surrogate model scales as $2\sigma^2 d_{\rm eff}/n$, the corresponding reduction in the generalization gap is of order $\Or(1/\sqrt{g})$. This reduction in effective complexity and generalization gap explains why moderate noise can improve testing performance.

\subsection{Mean-squared prediction error in the presence of quantum noise}

In this subsection, we derive the approximate training and testing MSE in the presence of quantum noise. For simplicity, we write
\begin{equation}
    s:=s_2(\beta,g)
\end{equation}
for the noise-order purity. We also use $\norm{\cdot}_2$ as the normalized empirical norm over the training inputs in this subsection.

Based on the local regression approximation in \cref{eqn:local-hat-matrix}, the surrogate model acts as a local smoother. Under the weak-overlap approximation of noise-order predictors, we have
\begin{equation}
    H(\boldsymbol{\theta};\beta,g)\cong s P_{\mc{T}},
\end{equation}
where $P_{\mc{T}}$ is the orthogonal projection onto the local learnable space $\mc{T}$, and $\mathrm{rank}(P_{\mc{T}})=q$. We decompose the true signal as
\begin{equation}
    f=f_{\parallel}+f_{\perp},\qquad f_{\parallel}=P_{\mc{T}}f,\quad f_{\perp}=(I-P_{\mc{T}})f.
\end{equation}
Here, $f_{\parallel}$ is the learnable component in the local model space, and $f_{\perp}$ is the component outside the local model space.

We assume homoscedastic data noise
\begin{equation}
    y=f+\epsilon,\qquad \mathbb{E}(\epsilon)=0,\qquad \mathbb{E}(\epsilon\epsilon^\top)=\sigma^2 I_n.
\end{equation}
The training MSE is then approximated by
\begin{equation}
\begin{split}
    \mathbb{E}\mathrm{MSE}_{\rm train}(\boldsymbol{\theta};\beta,g)&=\mathbb{E}\norm{\phi(\boldsymbol{\theta},\beta)-y}_2^2\approx \mathbb{E}\norm{(H(\boldsymbol{\theta};\beta,g)-I)y}_2^2\\
    &\approx \norm{f_{\perp}}_2^2+(1-s)^2\norm{f_{\parallel}}_2^2+\frac{\sigma^2}{n}(n-2qs+qs^2).
\end{split}
\end{equation}
The first term is the unlearnable component outside the local model space. The second term is the bias induced by noise. It increases when the noise-order purity $s$ decreases. The third term is the contribution from data noise. This expression shows that the training MSE increases as the noise rate increases.

Using \cref{eqn:generalization_gap,eqn:d_eff_s2}, the generalization gap is
\begin{equation}
    \Delta_{\rm gen}\approx \frac{2\sigma^2 q}{n}s.
\end{equation}
The testing MSE is the sum of the training MSE and the generalization gap. Adding this expression to the training MSE gives
\begin{equation}
    \mathbb{E}\mathrm{MSE}_{\rm test}(\boldsymbol{\theta};\beta,g)\approx \norm{f_{\perp}}_2^2+(1-s)^2\norm{f_{\parallel}}_2^2+\sigma^2+\frac{\sigma^2 q}{n}s^2.
\end{equation}
The generalization gap cancels the negative term $-2\sigma^2qs/n$ in the training MSE, which arises from fitting data noise in the training set.
As the noise rate increases, $s=s_2(\beta,g)$ decreases. This reduces the generalization gap, but it also increases the bias term $(1-s)^2\norm{f_{\parallel}}_2^2$. The competition between these two effects leads to a non-monotonic testing MSE.

The testing MSE is minimized at
\begin{equation}
    s^\star=\frac{\norm{f_{\parallel}}_2^2}{\norm{f_{\parallel}}_2^2+\sigma^2 q/n}=\frac{\mathrm{SNR}}{\mathrm{SNR}+q/n},
\end{equation}
where $\mathrm{SNR}:=\norm{f_{\parallel}}_2^2/\sigma^2$ is the signal-to-noise ratio.
The corresponding optimal noise rate $\beta^\star$ is determined by
\begin{equation}
    s_2(\beta^\star,g)=s^\star,
\end{equation}
when such a solution is in the achievable range of $s_2(\beta,g)$. This shows that noise is more likely to improve the testing MSE when the learnable signal is weak, the data noise is large, the locally learnable model dimension is large, or the training set is small.

As a remark, in the derivation above, we assume for simplicity that the model parameter count is smaller than the training data size. In this case, the rank of the projector $P_\mc{T}$ is simplified to $q$. When the model is overparameterized, however, the rank of the projector is no longer simply given by the model parameter count, but is jointly upper bounded by the model parameter count and the training data size. In this case, this rank should be replaced by $d_{\rm eff}^{(0)}$, the zero-noise effective model complexity. The corresponding form is
\begin{equation}
    s^\star\approx\frac{\mathrm{SNR}}{\mathrm{SNR}+d_{\rm eff}^{(0)}/n}.
\end{equation}

\subsection{Extensions beyond the Bernoulli noise model}

The analysis above uses the simple stochastic noise model in \cref{eqn:stochastic-noise-model} for clarity. In this model, each noisy location either applies the identity channel or applies a noisy operation. This gives a binomial expansion where the zero-noise branch corresponds to the noiseless implementation. However, this identity branch is not essential to the mechanism.

A direct extension is to replace the identity branch by a general principal channel. For example, suppose the channel at the $j$-th noisy location is given by
\begin{equation}
\mc{E}_{j,z_j}=(1-z_j)\mc{A}_j+z_j\mc{B}_j,\quad \mathbb{E}_{z_j}(\mc{E}_{j,z_j})=(1-\beta)\mc{A}_j+\beta\mc{B}_j,
\end{equation}
where $z_j\sim\mathrm{Ber}(\beta)$, and $\mc{A}_j$ and $\mc{B}_j$ are two quantum channels. The same noise-order expansion still applies, but the principal branch is no longer the noiseless implementation. Instead, the principal branch corresponds to the gate sequence distorted by the channel sequence ${\mc{A}_j}$. This differs from the simple identity-branch case, where the noiseless model hypothesis space is preserved. In this more general setting, the hypothesis space itself is distorted by the principal noise branch. Nevertheless, the same binomial weights $p_m(\beta,g)$ and the same noise-order purity $s_2(\beta,g)$ appear in the analysis, although the detailed predictors $\psi_m$ are changed.

More general noise models may contain more than two branches. In that case, the noisy channel can be viewed as a mixture of several effective operations. A full noise realization is specified by a path of branch choices across the circuit. The binomial weights are then replaced by the corresponding path probabilities. We can similarly define a generalized noise-order purity by summing the squared probabilities of these paths. The same intuition remains: when the path probabilities spread over more effective noise realizations, the surrogate model has a stronger averaging effect; when the noise becomes too strong, the learnable signal can also be suppressed.

\section{Numerical validation of the noise-order surrogate}
\label{sec:appendix-surrogate}

\subsection{Surrogate effective complexity in trained models}
\label{sec:appendix-surrogate-results}

\cref{sec:appendix-A} predicts that averaging over noise orders reduces the local effective complexity of the stochastic surrogate model. The main text tests this prediction using the Jacobian of the physical noisy predictor in \cref{fig:actual-deff}. Here we evaluate the surrogate complexity directly from the order-resolved Jacobians $J_m$ defined in \cref{eqn:order-jacobians}, providing a separate numerical test of the approximation leading to \cref{eqn:d_eff_s2}. These Jacobians are evaluated efficiently using the dynamic-programming procedure in \cref{sec:appendix-surrogate-sectors}.

For a fixed trained checkpoint and noise rate, let $J=\sum_m p_m(\beta,g)J_m$ denote the Jacobian of the averaged noisy predictor and define
\begin{equation}
    A(\beta)=\frac{1}{n}\sum_{m=0}^{g}p_m(\beta,g)J_m^\top J_m,\qquad B(\beta)=\frac{1}{n}J^\top J .
\end{equation}
The local hat matrix in \cref{eqn:local-hat-matrix} then gives the numerical surrogate complexity
\begin{equation}
    d_{\rm eff}^{\rm sur}(\beta)=\Tr\left((A(\beta)+10^{-4}I_q)^{-1}B(\beta)\right).
    \label{eqn:surrogate-deff-numerical}
\end{equation}
The matrix $A(\beta)$ can be singular, so we use the same $10^{-4}$ soft-counting threshold as in \cref{eq:methods-deff} to set the spectral resolution of its inverse. This threshold is applied only to the post-training model complexity evaluation and does not modify the training objective. For each run, we normalize $d_{\rm eff}^{\rm sur}(\beta)$ by its value at $\beta=0$ before averaging over the ten trained runs.

\begin{figure}[t]
    \centering
    \includegraphics[width=0.95\linewidth]{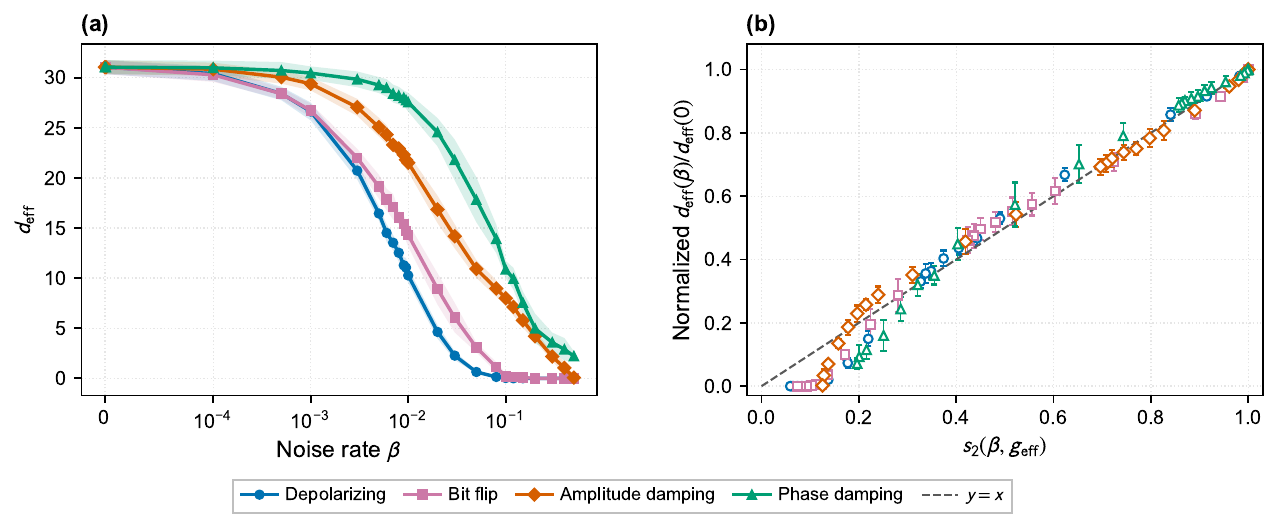}
    \caption{Effective model complexity of the noise-order surrogate. (a) Surrogate effective model complexity as a function of noise rate. Lines show the mean over ten runs and shaded regions show one standard deviation. (b) Effective model complexity normalized by its zero-noise value and plotted against $s_2(\beta,g_{\rm eff})$. Error bars show one standard deviation; fitted values of $g_{\rm eff}$ are reported in the text.}
    \label{fig:surrogate-deff}
\end{figure}

The surrogate complexity decreases with noise for all four noise models in \cref{fig:surrogate-deff}(a). After normalization, the noise-model-dependent curves are described by the same noise-order purity $s_2(\beta,g_{\rm eff})$ in \cref{fig:surrogate-deff}(b). The fitted effective counts are summarized in \cref{tab:surrogate-geff}. The surrogate loss explicitly penalizes disagreement among noise-order predictors, whereas the actual loss captures this regularization only implicitly through the structure of their averaged prediction. The same physical noise therefore induces stronger effective regularization in the surrogate, as reflected by its larger fitted $g_{\rm eff}$. Thus, the surrogate calculation reproduces the same $s_2$ scaling predicted by \cref{eqn:d_eff_s2} and observed in \cref{fig:actual-deff} for the physical-model complexity computed from the actual loss.

The surrogate makes the noise-order regularization explicit by retaining the variation among order-resolved Jacobians in its local curvature, whereas the physical-model complexity captures its consequence through the Jacobian of the averaged noisy predictor. Indeed,
\begin{equation}
    A(\beta)-B(\beta)=\frac{1}{n}\sum_{m=0}^{g}p_m(\beta,g)(J_m-J)^\top(J_m-J)\succeq0.
    \label{eqn:surrogate-curvature-gap}
\end{equation}
This matrix order inequality implies that the surrogate model complexity is no greater than the physical-model complexity. This confirms that the surrogate model makes the noise-order regularization explicit.

\subsection{Interpretation of the effective noise count}
\label{sec:appendix-surrogate-geff}

As shown in \cref{tab:surrogate-geff}, both calculations give the ordering
\begin{equation}
    g_{\rm eff}^{\rm depolarizing}>g_{\rm eff}^{\rm bit\ flip}>g_{\rm eff}^{\rm amplitude}>g_{\rm eff}^{\rm phase}.
\end{equation}
\begin{table}[H]
    \centering
    \caption{Fitted effective noise counts obtained from the ten-run mean normalized complexity curves in \cref{fig:actual-deff,fig:surrogate-deff}. Both calculations use the soft-counting threshold $10^{-4}$.}
    \label{tab:surrogate-geff}
    \setlength{\tabcolsep}{9pt}
    \begin{tabular}{lcccc}
        \toprule
         & Depolarizing & Bit flip & Amplitude damping & Phase damping \\
        \midrule
        Physical model & $32$ & $24$ & $11$ & $6$ \\
        Surrogate & $91$ & $59$ & $20$ & $8$ \\
        \bottomrule
    \end{tabular}
\end{table}
This ordering can be understood from how each noise model changes components of the state that can reach the final $Z^{\otimes4}$ measurement. Phase damping suppresses off-diagonal coherences but leaves computational-basis populations unchanged at the location where it acts. It affects the measured output only when subsequent gates convert the lost coherence into population differences, so many physical phase-damping locations have a weak effect on the predictor Jacobian. This gives the smallest $g_{\rm eff}$.

Amplitude damping also suppresses coherence, but it additionally transfers population from $\lvert1\rangle$ to $\lvert0\rangle$. It can therefore change a later $Z$ measurement more directly than phase damping. Bit-flip noise exchanges the two computational-basis populations and reverses the sign of the local $Z$ component, producing a still stronger response in the measured parity. Depolarizing noise contracts every traceless single-qubit Pauli component and therefore perturbs the broadest set of population and coherence directions that can contribute to the final prediction. It consequently gives the largest effective count.

Earlier errors can be rotated between population and coherence components by the remaining circuit, so their effects depend on the trained parameters, input states, and measured observable. The fitted $g_{\rm eff}$ summarizes the accumulated effect of each noise model on the measured predictor Jacobian.

\subsection{Efficient evaluation of noisy order-resolved quantities by dynamic programming}
\label{sec:appendix-surrogate-sectors}

At each of the $g$ noisy locations, the physical channel at a fixed $\beta$ is written exactly as
\begin{equation}
    \mc{E}_{\beta,\ell}=(1-\beta)\mathrm{Id}+\beta\mc{N}_{\beta,\ell},\qquad \mc{N}_{\beta,\ell}(\rho)=\frac{\mc{E}_{\beta,\ell}(\rho)-(1-\beta)\rho}{\beta}.
    \label{eqn:event-map-split}
\end{equation}
When evaluating the order-resolved Jacobians in \cref{sec:appendix-surrogate-results}, expanding this decomposition across $g$ noisy locations produces $2^g$ event placements, with $\binom{g}{m}$ placements at noise order $m$. Direct evaluation is therefore exponential in $g$. We instead use dynamic programming to accumulate all placements of the same order in a single averaged state.

To derive the recurrence, let $S\subseteq\{1,\ldots,\ell\}$ denote the event locations among the first $\ell$ noisy locations, and let $\rho_S$ be the state generated by that placement. We define the order-$m$ state as the conditional average over all placements of that order,
\begin{equation}
    \rho_m^{(\ell)}=\frac{1}{\binom{\ell}{m}}\sum_{\substack{S\subseteq\{1,\ldots,\ell\}\\ |S|=m}}\rho_S .
    \label{eqn:conditional-order-state}
\end{equation}
When location $\ell$ is added, an order-$m$ placement either retains order $m$ from the first $\ell-1$ locations or adds an event to an order-$(m-1)$ placement. Their multiplicities are $\binom{\ell-1}{m}$ and $\binom{\ell-1}{m-1}$, respectively. A one-step analysis therefore gives the recurrence
\begin{equation}
    \rho_m^{(\ell)}=\frac{\ell-m}{\ell}\rho_m^{(\ell-1)}+\frac{m}{\ell}\mc{N}_{\beta,\ell}\left(\rho_{m-1}^{(\ell-1)}\right).
    \label{eqn:noise-order-dp}
\end{equation}
Before this update, every retained sector is propagated through the circuit gates between locations $\ell-1$ and $\ell$. Since all unitary, channel, and residual maps are linear, averaging placements before subsequent propagation gives exactly the same result as propagating every placement separately and averaging at the end. \cref{eqn:noise-order-dp} therefore includes all $\binom{\ell}{m}$ possible locations at a given order without assuming that different circuit locations are equivalent.

By linearity, after all $g$ locations, the physical state and prediction are reconstructed as
\begin{equation}
    \rho(\beta)=\sum_{m=0}^{g}p_m(\beta,g)\rho_m^{(g)},\qquad \phi_\beta=\sum_{m=0}^{g}p_m(\beta,g)\psi_m,\qquad \psi_m=\Tr\left(O\rho_m^{(g)}\right).
    \label{eqn:order-reconstruction}
\end{equation}
No additional combinatorial factor appears in this expression: $\rho_m^{(g)}$ is already the conditional average within the sector, while $p_m(\beta,g)$ is the total probability mass of that sector.

The order-resolved Jacobians are obtained in the same forward pass by propagating the tangents $T_{m,j}^{(\ell)}=\partial\rho_m^{(\ell)}/\partial\theta_j$ alongside each state sector. Because the channel maps are linear, the tangents obey the same order recurrence in \cref{eqn:noise-order-dp}; parametrized gates contribute through their analytic derivatives. Measuring the final tangents gives $J_m$, whose weighted sum recovers the full Jacobian $J$ in \cref{eqn:order-jacobians}.

\begin{breakablealgorithm}
\caption{Order-resolved state and Jacobian propagation}
\label{alg:noise-order-dp}
\begin{algorithmic}[1]
\Require Input state $\rho_0$, parameters $\boldsymbol{\theta}$, and event maps $\mc{N}_{\beta,\ell}$
\State $\rho[0]\gets\rho_0$ and $T[0,j]\gets0$ for $j=1,\ldots,q$
\State Treat entries outside the current order range as zero
\State $\ell\gets0$
\For{each circuit gate $U$ in execution order}
    \State Propagate every $\rho[m]$ and $T[m,j]$ through $U$
    \If{$U=U_j(\theta_j)$ is parametrized}
        \State Add the analytic gate-derivative contribution to $T[m,j]$ for every $m$
    \EndIf
    \For{each noisy location following $U$}
        \State $\ell\gets\ell+1$ and copy the current arrays to $\rho_{\rm old}$ and $T_{\rm old}$
        \For{$m=0,\ldots,\ell$}
            \State Update $\rho[m]$ from $\rho_{\rm old}[m]$ and $\mc{N}_{\beta,\ell}(\rho_{\rm old}[m-1])$ using \cref{eqn:noise-order-dp}
            \State Apply the same recurrence to $T[m,j]$ for all $j$
        \EndFor
    \EndFor
\EndFor
\State Measure $\rho[m]$ and $T[m,j]$ to obtain $\psi_m$ and $J_m$
\State \Return $\{\psi_m,J_m\}_{m=0}^{g}$
\end{algorithmic}
\end{breakablealgorithm}

Let $n$ be the number of inputs, $q$ the number of trainable parameters, and $d=2^{N_{\rm q}}$ the Hilbert-space dimension. Retaining all noise orders requires $\Or(nqg^2d^3)$ time because the algorithm propagates at most $g+1$ sectors and their $q$ tangents through $g$ noisy locations. This is polynomial in $g$, in contrast to the $\Or(2^g)$ cost of enumerating all noise placements. In numerical calculations, sectors with negligible binomial probability can be omitted to further reduce the computational cost.

\section{Molecular QML simulation details and ablation studies}
\label{sec:appendix-molecular-details}

\subsection{Circuit structure and other simulation details}
\label{sec:appendix-molecular-circuit}

The molecular models use one qubit for each heavy atom and one EDU-QGC-inspired graph-circuit layer~\cite{Ceylan21_211205261,Rhee23_4300,Li26_260113275}. The atom encoder applies trainable element-dependent rotations modulated by the atom degree, hydrogen count and aromaticity, together with a shared rotation that encodes hybridization. A trainable three-angle rotation is then applied to each node. For every molecular bond, the circuit applies a bond-type-dependent local basis change, a CNOT--$R_Z$--CNOT interaction and a phase rotation, followed by the inverse basis change. Parameters in the bond block are shared across bonds of the same type.

The eight-qubit circuit contains 53 trainable quantum parameters, and the nine-qubit circuit contains 56. The diagonal of the final density matrix gives 256 and 512 computational-basis probabilities, respectively. A linear map from these probabilities to the molecular target adds 257 parameters for QM9-HA8 and 513 for PCQM4Mv2-HA9, giving 310 and 569 trainable parameters in the complete predictors.

\label{sec:appendix-molecular-noise-count}

A noisy location is assigned to each qubit immediately after an elementary gate acts on it. A single-qubit gate therefore contributes one location, while a two-qubit gate contributes two. In the four-qubit circuit, each layer contains two data-encoding locations, eight locations from the four two-qubit gates and four locations from the trainable single-qubit gates. The nine layers therefore give
\begin{equation}
    g=9(2+8+4)=126.
\end{equation}

For the molecular models, the atom encoders and node rotations contribute locations on every atom, while the bond blocks contribute locations along the chemical edges. The resulting count $g(G)$ therefore depends on the molecular graph $G$. The physical noisy simulation applies the channel at every executed location, while the fitted $g_{\rm eff}$ summarizes how rapidly the effective model complexity decreases with noise across the molecular dataset.

\subsection{Ablation of optimization epochs}
\label{sec:appendix-molecular-optimization}

\begin{figure}[t]
    \centering
    \includegraphics[width=0.99\linewidth]{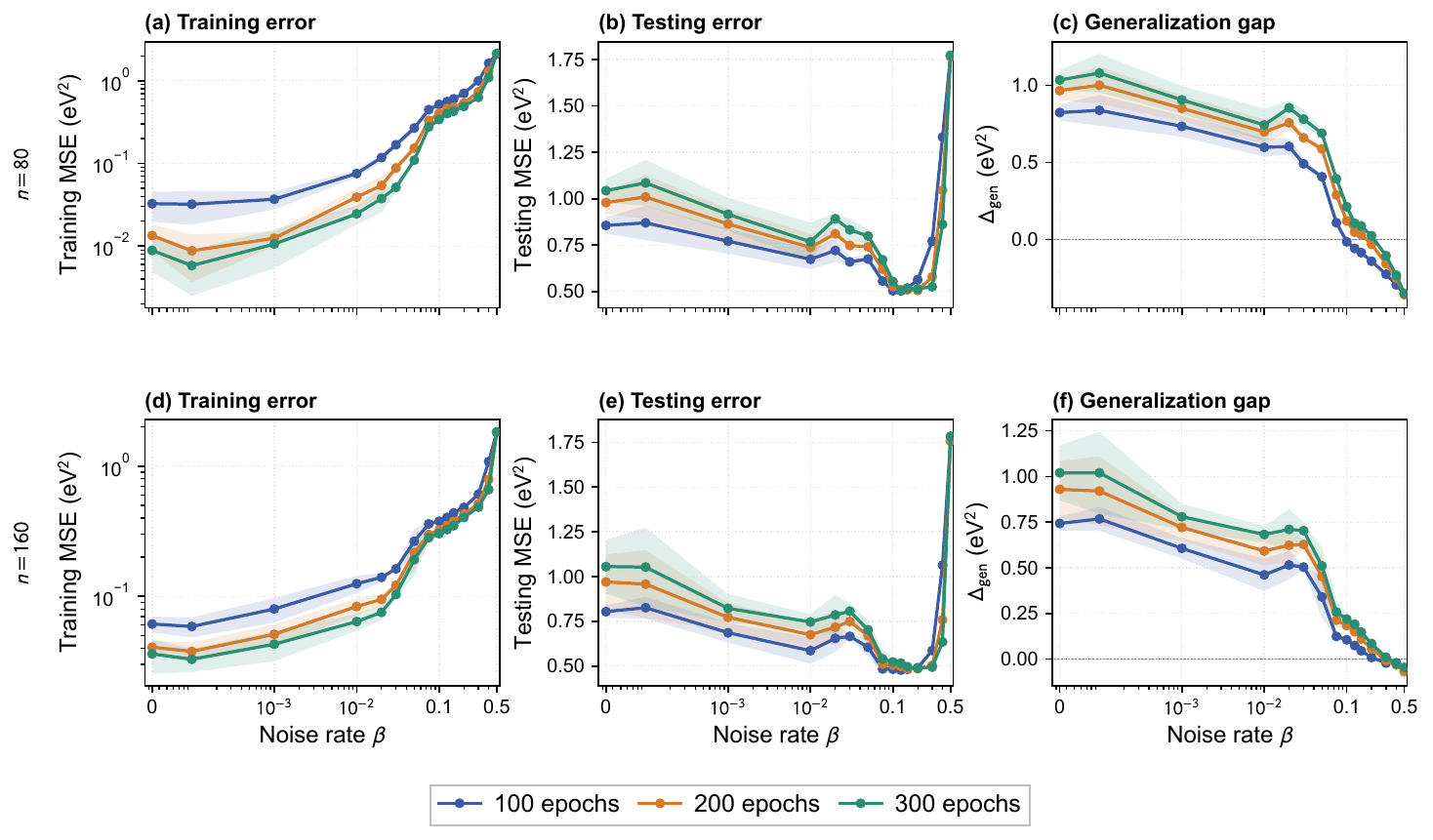}
    \caption{Ablation of optimization epochs. Training MSE, testing MSE and generalization gap are evaluated after 100, 200 and 300 epochs for QM9-HA8 with (a--c) $n=80$ and (d--f) $n=160$ training molecules. Lines show the mean over five fixed runs, and shaded regions show one sample standard deviation.}
    \label{fig:appendix-molecular-optimization}
\end{figure}

We evaluate the same five QM9-HA8 training trajectories after 100, 200 and 300 epochs to determine whether the finite-noise response depends on the optimization endpoint. With longer training, the training error decreases while the testing error and generalization gap at low noise increase, indicating stronger overfitting (\cref{fig:appendix-molecular-optimization}). The overall non-monotonic response remains stable across the three optimization endpoints.

\subsection{Ablation of more complex classical post-processing}
\label{sec:appendix-molecular-readout}

\begin{figure}[t]
    \centering
    \includegraphics[width=0.99\linewidth]{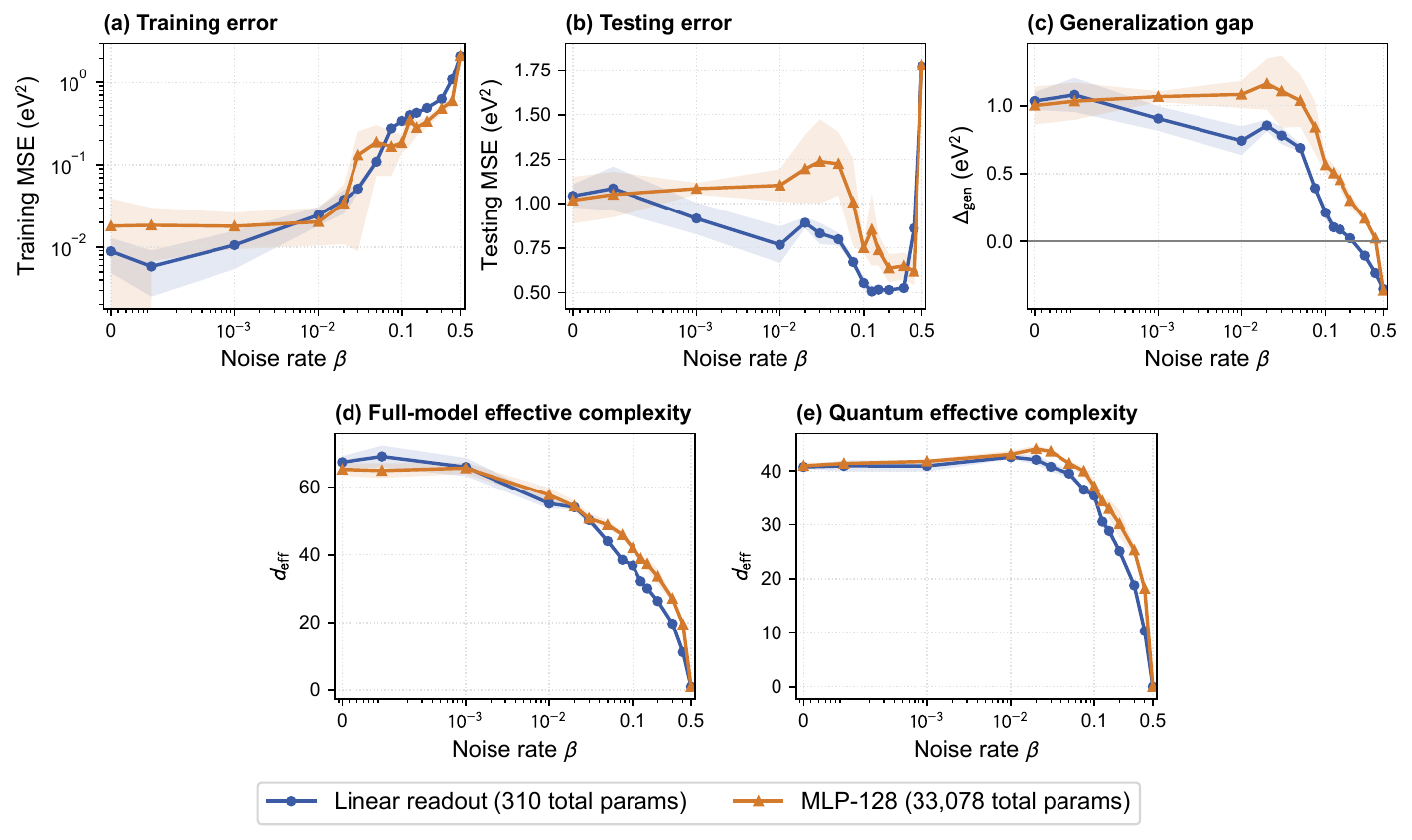}
    \caption{Ablation of more complex classical post-processing. (a--c) Training MSE, testing MSE and generalization gap for a linear readout and a nonlinear readout with 128 tanh hidden units. (d,e) Absolute effective model complexities computed using the full parameter vector and the quantum-circuit parameters only, respectively. Both models use the same 53-parameter quantum circuit, $n=80$ QM9-HA8 training molecules, bit-flip noise and 300-epoch endpoint. Lines and shaded regions show the mean and one standard deviation over five fixed runs.}
    \label{fig:appendix-molecular-readout}
\end{figure}

The molecular QML model uses a trainable classical map from the computational-basis probabilities to the molecular target. To test whether the finite-noise improvement depends on the linear readout, we replace it with a fully connected hidden layer containing 128 tanh units while keeping the quantum circuit, training set and noise implementation fixed. This increases the total parameter count from 310 to 33,078. Both readouts fit the low-noise training data closely and retain a lower testing error at finite noise (\cref{fig:appendix-molecular-readout}(a--c)). The nonlinear readout produces a more variable response across the noise sweep but does not improve the minimum testing error. The finite-noise improvement is therefore not specific to the choice of classical post-processing, and additional classical capacity alone does not improve prediction.

This behavior is also reflected in the effective model complexity (\cref{fig:appendix-molecular-readout}(d,e)). Despite their very different classical parameter counts, the two models have comparable full-model and quantum-only complexities across the noise sweep. The reduction in model complexity is therefore governed primarily by the noisy quantum representation rather than the added classical capacity.

\subsection{Ablation of noise models}
\label{sec:appendix-molecular-noise-models}

\begin{figure}[t]
    \centering
    \includegraphics[width=0.99\linewidth]{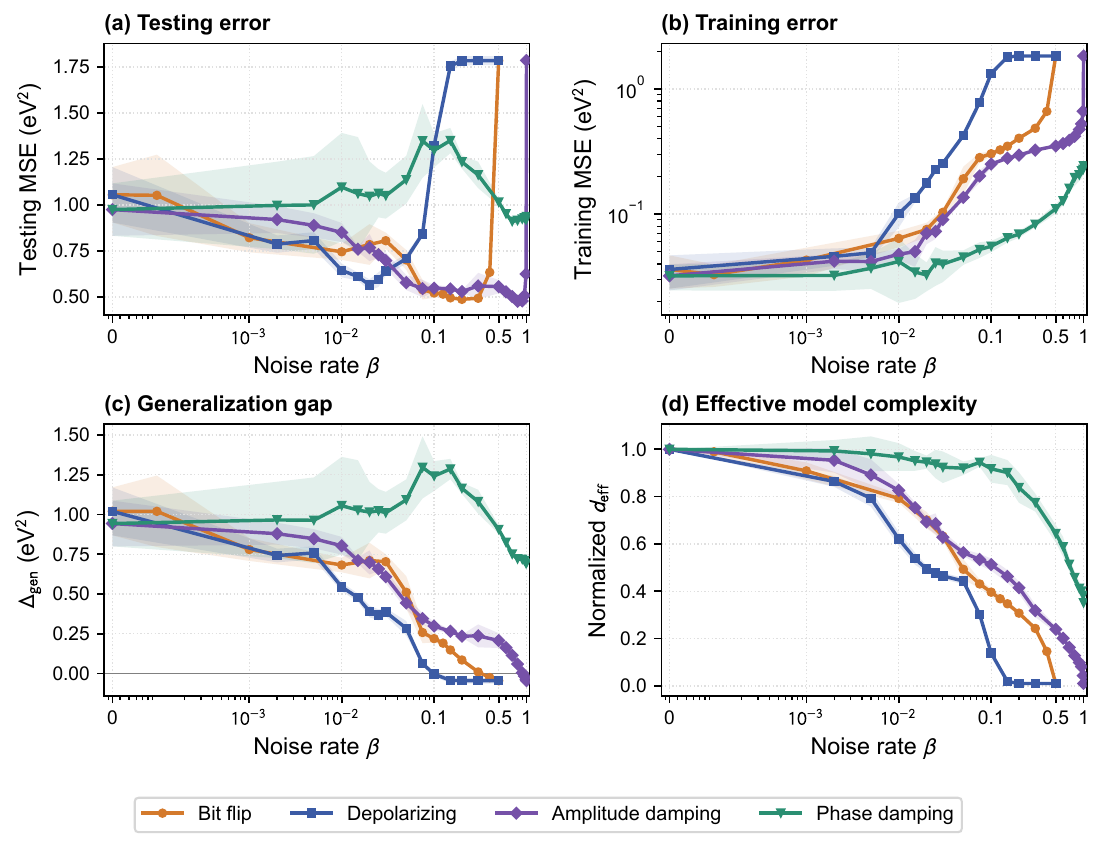}
    \caption{Ablation of noise models in the molecular QML model. (a) Testing MSE, (b) training MSE, (c) generalization gap and (d) effective model complexity normalized by its zero-noise value for bit-flip, depolarizing, amplitude-damping and phase-damping noise. The damping channels are shown over $0\leq\beta\leq1$, while bit-flip and depolarizing noise are shown over $0\leq\beta\leq0.5$. All models use the same QM9-HA8 training set with $n=160$ and are evaluated after 300 epochs. Lines and shaded regions show the mean and one standard deviation over five fixed runs.}
    \label{fig:appendix-molecular-noise-models}
\end{figure}

The main molecular comparison uses bit-flip and depolarizing noise, while the full comparison across noise models is reported here without crowding the main figure. We compare all four noise models for the same QM9-HA8 predictor, training set and optimization settings. Bit flip, depolarizing and amplitude damping produce a clear intermediate-noise reduction in testing error, whereas phase damping does not produce the same non-monotonic response (\cref{fig:appendix-molecular-noise-models}(a)). The corresponding training errors show that these minima occur before strong noise removes the ability to fit the training data (\cref{fig:appendix-molecular-noise-models}(b)).

All four noise models reduce the effective model complexity, but at different rates (\cref{fig:appendix-molecular-noise-models}(d)). This difference explains why their minima occur at different physical noise rates: the same value of $\beta$ does not produce the same reduction in model complexity across noise models. The fitted $g_{\rm eff}$ accounts for this mapping in \cref{fig:molecular-qgnn}(f), where the normalized complexity curves are compared against the noise-order purity $s_2(\beta,g_{\rm eff})$. Thus, the channel dependence changes how quickly the model moves through the finite-noise regime, while its effect on prediction also depends on the measured representation.

\begin{figure}[t]
    \centering
    \includegraphics[width=0.99\linewidth]{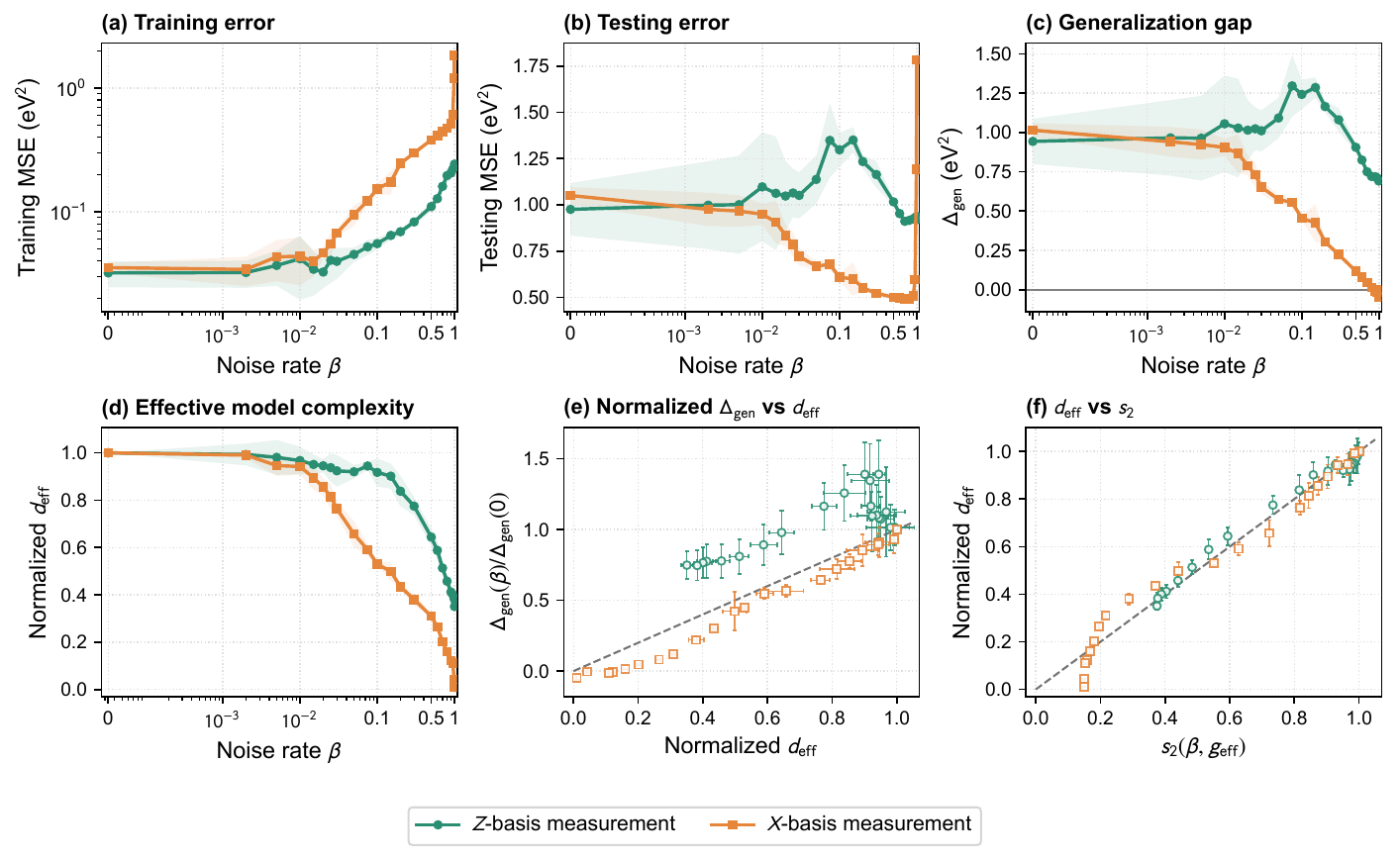}
    \caption{Measurement dependence under phase damping. (a) Training MSE, (b) testing MSE, (c) generalization gap and (d) effective model complexity normalized by its zero-noise value for full-probability measurements in the $Z$ and $X$ bases. (e) Normalized generalization gap versus normalized effective model complexity. (f) Normalized effective model complexity versus $s_2(\beta,g_{\rm eff})$, with $g_{\rm eff}$ fitted separately for the two bases. Green circles and orange squares denote the $Z$- and $X$-basis measurements, respectively. Both models use the same QM9-HA8 training set with $n=160$ and are evaluated after 300 epochs. Lines and shaded regions in (a--d), and points and error bars in (e,f), show the mean and one standard deviation over five fixed runs.}
    \label{fig:appendix-molecular-phase-readout}
\end{figure}

Under phase damping, the training error increases with noise, but the generalization gap does not decrease monotonically. The testing error therefore does not show the non-monotonic response observed for the other noise models. To test whether this behavior originates from the measurement basis, we repeat the phase-damping sweep using the same full-probability readout in either the $Z$ or $X$ basis (\cref{fig:appendix-molecular-phase-readout}). Switching to the $X$ basis recovers both the high-noise collapse and the linear relation between the generalization gap and effective model complexity, whereas the $Z$-basis predictor retains a nonzero complexity and deviates from this relation. The complexity itself remains well described by the noise-order purity in both bases. Phase damping removes coherence in the $Z$ basis, which directly affects an $X$-basis measurement but has a weaker effect on a $Z$-basis measurement. The benefit of noise-induced regularization therefore depends not only on the noise model, but also on the measured representation.

\bibliographystyle{unsrt}
\bibliography{ref}

\end{document}